\documentclass[aps,superscriptaddress]{revtex4-2}
\usepackage{floatrow}
\usepackage{epsfig,wrapfig}
\usepackage{amssymb,amsfonts,amsmath}
\usepackage{afterpage}
\usepackage{soul}

\usepackage{bm}% bold math
\usepackage{siunitx}
\usepackage{xcolor}

\usepackage[T1]{fontenc}
\usepackage[utf8]{inputenc}
\usepackage{babel}

\usepackage{hyperref} 

\newcommand{\be}{\begin{equation}}
\newcommand{\ee}{\end{equation}}
\newcommand{\bea}{\begin{eqnarray}}
\newcommand{\eea}{\end{eqnarray}}

\newcommand{\lc}{\color{black}}
\newcommand{\corr}{\color{black}} 
\newcommand{\corrprev}{\color{black}} 
\newcommand{\moveprev}{\color{black}}

\newcommand{\Ecoli}{{\it Escherichia coli}}
\newcommand{\ecoli}{{{\it E.~coli}}}
\newcommand{\Bsub}{{\it Bacillus subtilis}}

\newcommand{\Ccres}{{\it Caulobacter crescentus}}
\newcommand{\ccres}{{\it C.~crescentus}}
\newcommand{\Pseudo}{{\it Pseudomonas aeruginosa}}

\newcommand{\Lk}{\text{Lk}}
\newcommand{\Tw}{\text{Tw}}
\newcommand{\Wr}{\text{Wr}}
\newcommand{\s}{\sigma}

\begin{document}

\title{DNA supercoiling in bacteria: state of play and challenges from a {\corrprev viewpoint of physics based  modeling}}
\author{Ivan Junier}
\email{ivan.junier@univ-grenoble-alpes.fr}
\affiliation{Univ. Grenoble Alpes, CNRS, UMR 5525, VetAgro Sup, Grenoble INP, TIMC, 38000 Grenoble, France}
\author{Elham Ghobadpour}
\affiliation{Univ. Grenoble Alpes, CNRS, UMR 5525, VetAgro Sup, Grenoble INP, TIMC, 38000 Grenoble, France}
\affiliation{Université de Lyon, École Normale Supérieure (ENS) de Lyon, CNRS, Laboratoire de Physique and Centre Blaise Pascal de l’ENS de Lyon, F-69342 Lyon, France}
\author{Olivier Espeli} 
\affiliation{Center for Interdisciplinary Research in Biology (CIRB), Coll\`ege de France, CNRS, INSERM, Universit\'e PSL, Paris, France}
\author{Ralf Everaers}
\affiliation{Université de Lyon, École Normale Supérieure (ENS) de Lyon, CNRS, Laboratoire de Physique and Centre Blaise Pascal de l’ENS de Lyon, F-69342 Lyon, France}


\begin{abstract} 
DNA supercoiling is central to {\lc many} fundamental processes of living organisms. Its average level along the chromosome and over time reflects the dynamic equilibrium of opposite activities of topoisomerases, which are required to relax mechanical stresses that are inevitably produced during DNA replication and gene transcription. Supercoiling affects all scales of the spatio-temporal organization of bacterial DNA, from the base pair to the large scale chromosome conformation. Highlighted {\it in vitro} and {\it in vivo} in the 1960s and 1970s, respectively, the first physical models were proposed concomitantly in order to predict the deformation properties of the double helix. About fifteen years later, polymer physics models demonstrated on larger scales the plectonemic nature and the tree-like organization of supercoiled DNA. Since then, many works have tried to establish a better understanding of the multiple structuring and physiological properties of bacterial DNA in thermodynamic equilibrium and {\corrprev far from} equilibrium.

The purpose of this essay is to address upcoming challenges by {\corr thoroughly exploring} the relevance, predictive capacity, and limitations of current physical models, with a specific focus on structural properties beyond the scale of the double helix. {\corr We discuss more particularly the problem of DNA conformations, the interplay between DNA supercoiling with gene transcription and DNA replication, its role on nucleoid formation and, finally, the problem of scaling up models}. {\corr Our primary objective is to foster increased collaboration between physicists and biologists. To achieve this,} we have reduced the respective jargon to a minimum and we provide some explanatory background material for the two communities.
%{\corr In the following sections, we first discuss equilibrium properties of supercoiled DNA molecules as those measured in single-molecule experiments, and we next delve into the modeling of major biological processes occurring {\it in vivo}, including gene transcription, DNA replication, and nucleoid formation. Lastly, we address the challenging task of scaling up models.}
\end{abstract}

\maketitle 

\newpage
\tableofcontents
\newpage

With respect to DNA, efficient growth and division of bacteria rely on two major processes: (i) an appropriate expression of the genetic program allowing the generation in the right amounts and proportions of the proteins and enzymes necessary for the duplication of cells; (ii) a faithful replication of DNA and a reliable segregation of the replicated chromosomes during cell division. %A large part of modeling issues of the bacterial genome therefore concern the relationship between DNA structure and gene expression, on the one hand, and mechanisms associated with the correct replication and cellular distribution of chromosomes, on the other hand.
Research over the last fifty years or so has shown that the analysis of the topological constraints inherent in the double-helix nature of DNA is crucial for a quantitative understanding of these problems{\corrprev~\cite{wang_dna_1983,travers_dna_2005,dorman_dna_2016}}. Topological constraints are more particularly responsible for the supercoiling of bacterial DNA, {\corrprev i.e., the under or overwinding of bacterial DNA, which is} known to impact all levels of chromosome structure~\cite{wang_dna_1983,travers_dna_2005,badrinarayanan_bacterial_2015,dorman_dna_2016,dame_chromosome_2020,lioy_multiscale_2021}.
%, from the base pair to its large scale conformation 

%{\corr Research in DNA supercoiling has been involving different communities, including mathematicians, experimental and theoretical physicists, as well as various fields of biology, including molecular, cellular, and chromosome biologists.} 

%these biological issues from the perspective of physical modeling. %To this end, we review supercoiling-related} models that have been developed {\corrprev over the years} to explain {\it in vivo} properties of bacterial DNA occurring above the double helix scale.

Just as most fields of biology, investigation in {\corr the field of DNA supercoiling} has recently thrived thanks to a dramatic acceleration in the production of experimental results as a result of low-cost DNA sequencing, new genome engineering techniques and the development of visualization methods of increasing resolution. One of the consequences of having access to comprehensive data, some of which, such as high-throughput chromosome conformation capture (Hi-C) data~\cite{lieberman-aiden_comprehensive_2009}, covers almost all scales of a chromosome~\cite{lieberman-aiden_comprehensive_2009,le_new_2014}, is the possibility of building models of chromosomal organization {\corr across multiple genomic scales}. {\corr In this regard, it is essential to consider that the term {\it model} can have different meanings depending on the scientists' background, including biologists, modelers, and those with or without a physical background. For instance, in the context of chromosome structuring, {\it data-driven models}~\cite{rosa_computational_2014,junier_demultiplexing_2015,imakaev_modeling_2015} involve many parameters that may not be associated with any physical mechanism but, instead, used to generate, within a given polymer framework, chromosome conformations that are compatible with genome-wide data~\cite{umbarger_three-dimensional_2011,zhang_topology_2015,messelink_learning_2021} -- generated conformations can then be used to explore the statistical properties that underlie experimental data~\cite{zhang_topology_2015,messelink_learning_2021}}. On the other hand, {\it physics-based models} involve a set of {\corr physically motivated parameters, often parsimonious, and are used to {\it rationalize} observed experimental} data
%which involve idealized, {\corr often parsimonious}  mathematical representations of a {\corrprev specific} reality (physical, chemical and/or biological). {\corrprev These models, which incorporate} 
within the framework of the fundamental laws of Physics, particularly within the realm of Statistical Mechanics. In the case of DNA, the employed models often come from the neighboring fields of polymer physics and of soft and active matter~\cite{marko_biophysics_2015}.

In this review, we aim to discuss the problem of {\corr DNA supercoiling from this perspective of physical modeling, examining the components of biophysical models, their outcomes, as well as their limitations and possible workarounds. By doing so, we aim %not only to establish a common ground for stimulating further research at the interface between biophysics and microbiology but also
to clarify the open problems in the field, following the line of the famous quote by Richard Feynman: ``What I cannot create, I do not understand.'' To this end, we have divided} the review into {\corrprev seven} sections {\corrprev plus an \hyperref[sec:app_models]{Appendix}}. Section~\ref{sec:basics} revisits essential notions of DNA topology, introduces the molecular machines central to the problem, and discusses the problem of {\it in vivo} measurements {\corrprev of DNA supercoiling}. {\corrprev In section~\ref{sec:models}, we introduce the modeling approaches employed by biophysicists to comprehend and predict the behavior of supercoiled DNA, with additional details provided in the \hyperref[sec:app_models]{Appendix}.} Section~\ref{sec:equilibrium} presents the main steps marking the development of models aiming at capturing the {\it equilibrium properties} of supercoiled DNA, along with a discussion of their relevance for {\it in vivo} situations. Sections~\ref{sec:transcription} and~\ref{sec:replication} focus on transcription and replication, respectively, emphasizing the necessity to build {\it {\corrprev far from} equilibrium models} that involve not only the transcription and replication machineries but also the action of topoisomerases. In section~\ref{sec:nucleoid}, we discuss the formation of the nucleoid, which is the membrane-free region of the bacterial cells where DNA is found. In the final section~\ref{sec:scaleup}, we review the attempts to model the structuring of bacterial chromosomes at the largest scales.

%\section{DNA supercoiling in bacteria: connecting the multiple scales of a chromosome}
\section{DNA supercoiling in bacteria: fundamentals}
\label{sec:basics}

{\corr DNA is a polymer made up of nucleotides, arranged in a double helix structure formed by two intertwined strands, known as Watson and Crick strands, which are held together by hydrogen bonds.} In its relaxed state, at typical physiological temperature and salt concentration, a DNA double helix contains approximately $10.5$ base pairs (B-DNA form). However, in mesophilic bacteria, i.e., in bacteria living under mild conditions of temperature, pressure and pH, the double helix is generally longer, containing more than $10.5$ base pairs. Bacterial DNA is therefore under torsional stress, with an average underwound or, equivalently, negatively supercoiled double helix. This section explores the reasons behind these observations, starting with the notion of the linking number, the role of topoisomerases in relaxing torsional stresses generated during gene transcription and DNA replication, and the challenges of measuring the supercoiling properties of bacterial chromosomes.

%{\corrprev The DNA of most bacteria exists in a circular form. This characteristic has specific implications at all levels of bacterial chromosome structuring, ranging from the base pair to the large-scale chromosome conformation. The various conformations the chromosome can adopt must indeed be consistent with the so-called {\it conservation of the linking number}. To commence this section, we thus provide an explanation of the ``linking number'' concept.
%which allows is necessary to understand why local alterations in DNA helicity often impact the overall folding of the chromosome. 
%Subsequently, we introduce topoisomerases, crucial enzymes that ensure proper replication and transcription, whose activity on DNA actually modifies the linking number. Finally, we address the challenge of measuring the linking number {\it in vivo}. Overall, this initial section aims to clarify the intricacies associated with the commonly asserted statement: ``Bacterial DNA is negatively supercoiled''.}

\begin{figure}[t]
\centering
\includegraphics[width=0.95\linewidth]{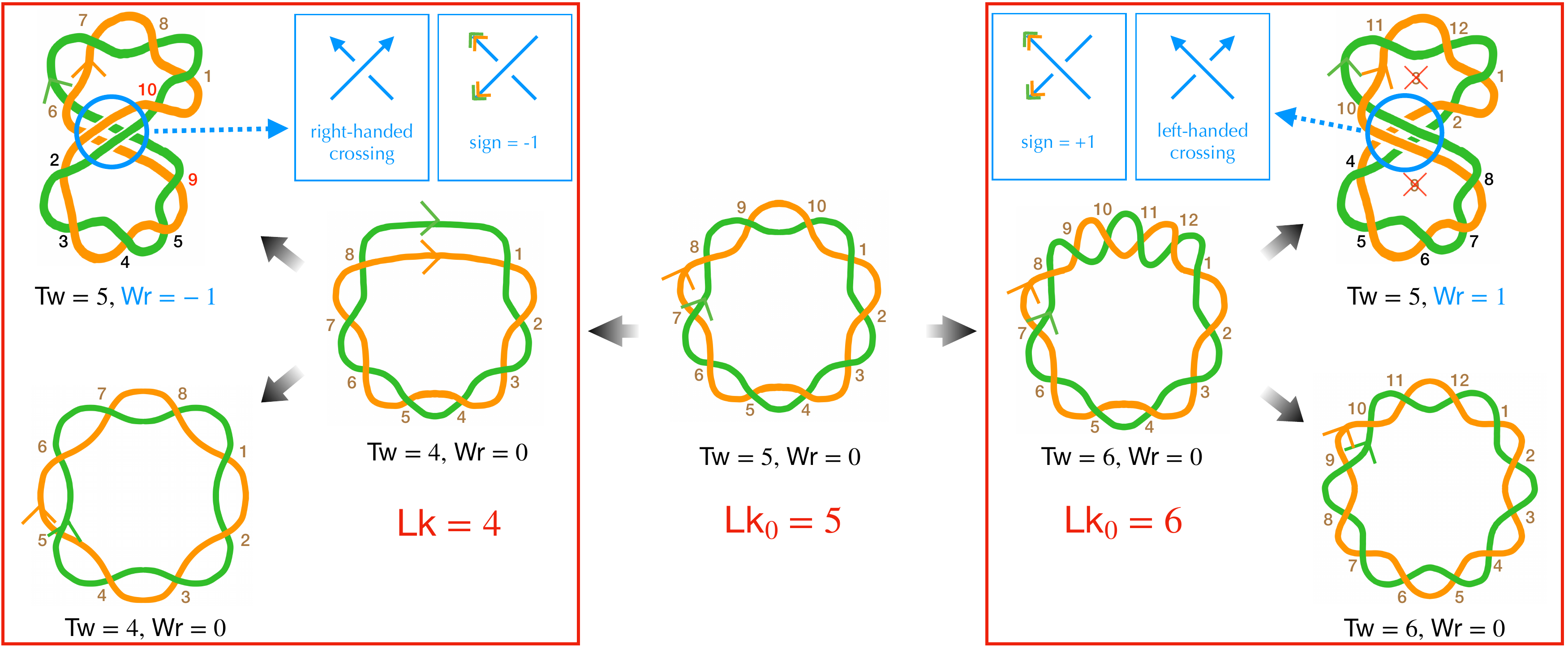}
\caption{{\lc Implications of the linking number conservation in circular DNA} {\corr -- only the DNA strands are depicted (in green and orange), presented schematically to emphasize topological properties}.
{\lc {\it Center}: we consider a reference molecule, torsionally relaxed and planar ($\Wr=0$), consisting of five helix turns ($\Tw=5$), resulting in a relaxed linking number $\Lk_0=5$. Brown numbers indicate right-handed DNA helix crossings. {\it Left}: we remove one helix turn (negative supercoiling), resulting in a molecule with heterogeneous helicity, comprised of four helix turns (right conformation), leading to $\Lk=4$ with $\Tw=4$ and $\Wr=0$. Note that the half-turn at the top of this conformation strongly distorts the double helix and likely denatures in real situations. One possibility is that the helix turns redistribute, achieving homogeneous helicity (bottom left conformation), the writhe and twist remaining unchanged. Alternatively, the molecule may buckle, forming a super-structure (top left conformation). In this case, the molecule can recover its relaxed twist ($\Tw=5$) if the super-structure is right-handed, with a corresponding $\Wr=-1$, by allowing the strands to cross two more times around the main axis, as indicated by the red numbers. In this conformation, the black numbers indicate helix crossings with a change in the strand passing on top of the other one, as a consequence of the buckling, the handedness of the helix remaining unchanged. {\it Right}: we introduce one helix turn (positive supercoiling). Qualitatively, the discussion resembles that of negative supercoiling, with one notable difference: to achieve the relaxed twist, a helix turn must be removed, not added. As indicated by the crossed numbers, this can occur with a left-handed super-structure, characterized by $\Wr=+1$. Finally, we remind that determining the handedness of the super-structure is based on the same rule as for the DNA double helix to indicate the directions of the main axis (blue arrows in the top inset panels). The sign of the corresponding writhe is instead determined using the directions as given by the DNA strands (orange and blue arrowheads in the top inset panels).}}
\label{fig:linking}
\end{figure}

\subsection{{\corrprev Linking number, twist/writhe decomposition and structural consequences}}

{\corrprev
The DNA of most bacteria exists in a circular form. This characteristic has specific implications at all levels of bacterial chromosome structuring, ranging from the base pair to the large-scale chromosome conformation. The various conformations the chromosome can adopt must indeed be consistent with the so-called {\it conservation of the linking number}. Specifically, the linking number ($\Lk$) of a circular DNA molecule represents the number of times the two DNA strands intersect in the three-dimensional space. For example, in a planar molecule, $\Lk$ is equal to the number of helix turns the two strands make along the molecule's central axis -- this can be calculated by considering that one helix turn of B-DNA consists of approximately $10.5$ base pairs. In the more general case of a three-dimensional molecule, the strand intersections can occur locally as the strands twist around each other along the molecule's central axis, as well as globally when the main axis folds and crosses itself (Fig.~\ref{fig:linking}). Consequently, $\Lk$ is the combined result of the twist ($\Tw$) and the writhe ($\Wr$), expressed as $\Lk = \Tw + \Wr$~\cite{calugareanu_lintegrale_1959, white_self-linking_1969,fuller_decomposition_1978}. The twist refers to the total number of helix turns, while the writhe represents the average number of times the main axis crosses itself from any perspective~\cite{fuller_decomposition_1978}.

{\it DNA supercoiling} occurs when the linking number deviates from that of the corresponding mechanically relaxed molecule. This can happen {\it in vivo} due to the activity of enzymes, as discussed below, or {\it in vitro} when the DNA is manipulated, for example, by magnetic tweezers~\cite{strick_stretching_2003}. Note that, conventionally, a positive contribution to the twist indicates a helix involving right-handed intersections, while a positive contribution to the writhe signifies a left-handed intersection in space. Conversely, negative contributions for twist and writhe indicate a left-handed double helix and a right-handed intersection, respectively (Fig.~\ref{fig:linking}).

Importantly, for any deformation of the DNA molecule in which the two strands are not cut, the linking number remains unchanged~\cite{calugareanu_lintegrale_1959, white_self-linking_1969}. This property, known as the {\it conservation of the linking number}, implies that twist can precisely convert into writhe, and {\it vice versa}, as depicted in Fig.~\ref{fig:linking} {\corr -- describing DNA as a ribbon can further help to apprehend this property~\cite{crick_linking_1976,bauer1980supercoiled}}. This fundamental characteristic enables supercoiled DNA to relieve local torsional stress by generating super-structures, such as plectonemes. The relative proportions of deformations in the double helix and formation of super-structures are then determined by the energy costs associated with torsion and bending mechanical properties of DNA. Physical models have extensively focused on predicting both these proportions and the resulting conformations, as explained in detail in sections~\ref{sec:models} and~\ref{sec:equilibrium}.}

\begin{figure}[t]
\centering
\includegraphics[width=0.65\linewidth]{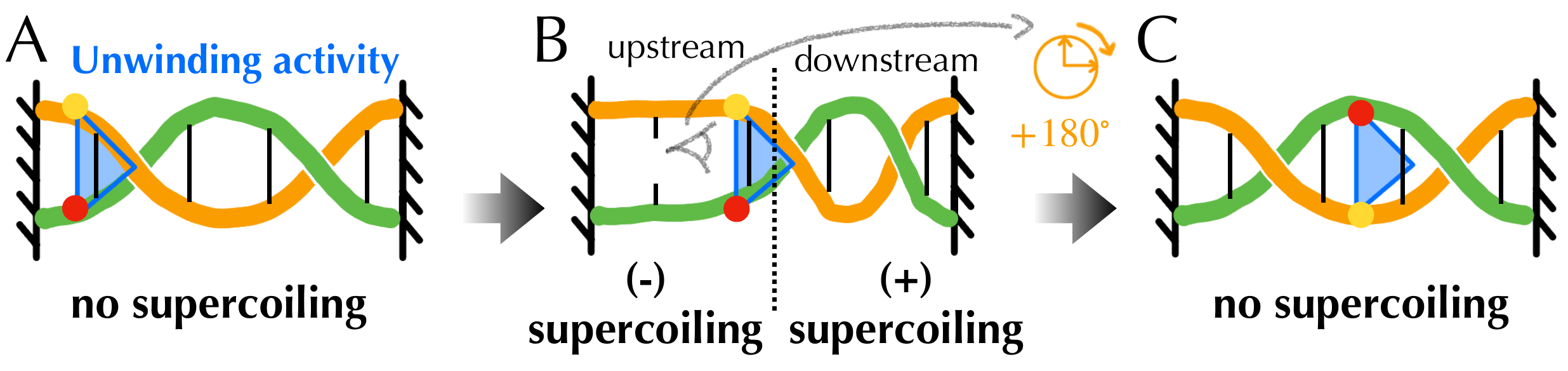}
\caption{Schematic representation of torsional stresses generated during translocation of a DNA unwinding machine, {\lc starting from a situation with no supercoiling.} 
Four base pairs (vertical lines) are indicated to facilitate reading. 
A) The DNA ends are prevented from rotating, mimicking the effect of a topological barrier. Two extreme possibilities can then be considered: B) If the unwinding machine does not rotate around the DNA, {\lc it behaves as a topological barrier and} the double helix becomes increasingly overwound downstream and underwound upstream, respectively generating positive and negative supercoiling. The latter can lead to DNA denaturation, as indicated by the breaking of the base pair. C) If the unwinding machine freely rotates around the DNA, the machine rotates clockwise while advancing along the undeformed right-handed DNA double helix.}
\label{fig:twin_scheme}
\end{figure}

\subsection{{\corrprev Topoisomerases: changing the DNA linking number {\corr when} resolving transient topological stresses}}
\label{sec:topo1}

{\corrprev 
When a circular B-DNA molecule is in its mechanically most relaxed state, 
%both the torsional and bending energies are close to their minimum. Correspondingly, 
the twist is close to the number of double helix turns, the writhe is negligible with respect to the twist, and the corresponding relaxed linking number, $\Lk_0$, is almost equal to the twist. However, {\it in vivo}, DNA {\lc undergoes} torsional stresses generated during DNA replication and gene transcription. These stresses are alleviated by DNA enzymes called topoisomerases~\cite{wang_cellular_2002,forterre_origin_2007,mckie_dna_2021}. By doing so, topoisomerases effectively change the overall linking number of DNA, leading to supercoiling of the bacterial chromosome wherein the linking number differs from $\Lk_0$.

Before providing details about topoisomerases, let us explicit the nature of the torsional stresses they relax during gene transcription and DNA replication. Namely, in both processes,} associated macromolecular complexes including the RNA and DNA polymerases locally open bacterial DNA and proceed along it in a specific direction. Multiple protein complexes are bound to this DNA, and the expected situation {\it in vivo} is that of a chromosome organized into DNA domains whose ends are prevented from rotating by topological barriers~\cite{liu_supercoiling_1987} (see section~\ref{sec:transcription} for details). Consider, in this case, a piece of DNA such that the Watson and Crick strands of the double helix are held in a rotationally fixed position at the borders (Fig.~\ref{fig:twin_scheme}A). Just as in a circular molecule, these constraints impose the conservation of the linking number between the two strands. Consider, then, an idealized machine locally opening the DNA and advancing along it (Fig.~\ref{fig:twin_scheme}BC).
To the extent that the local opening is associated with a local unwinding of the strands (not represented in Fig.~\ref{fig:twin_scheme} for clarity), the conservation of the linking number implies that the remaining double helical parts have to overwind in compensation. {\corrprev In this context}, the torsional stresses induced by the progressing machine depend on whether it can freely rotate around the DNA  (Fig.~\ref{fig:twin_scheme}C) or not (Fig.~\ref{fig:twin_scheme}B). 
In the former case, the machine rotates clockwise while advancing along the right-handed DNA double helix and no additional torsional stresses are exerted beyond those due to the initial opening. In the latter case, the double helix becomes increasingly overwound downstream and underwound upstream. This means that the number of base pairs per helix turn decreases or increases correspondingly. The progression thus induces respectively positive downstream and negative upstream {\it twin} DNA supercoiling~\cite{liu_supercoiling_1987}, although {\it no net overall supercoiling} has been introduced.

\begin{figure}[t]
\centering
\includegraphics[width=0.5\linewidth]{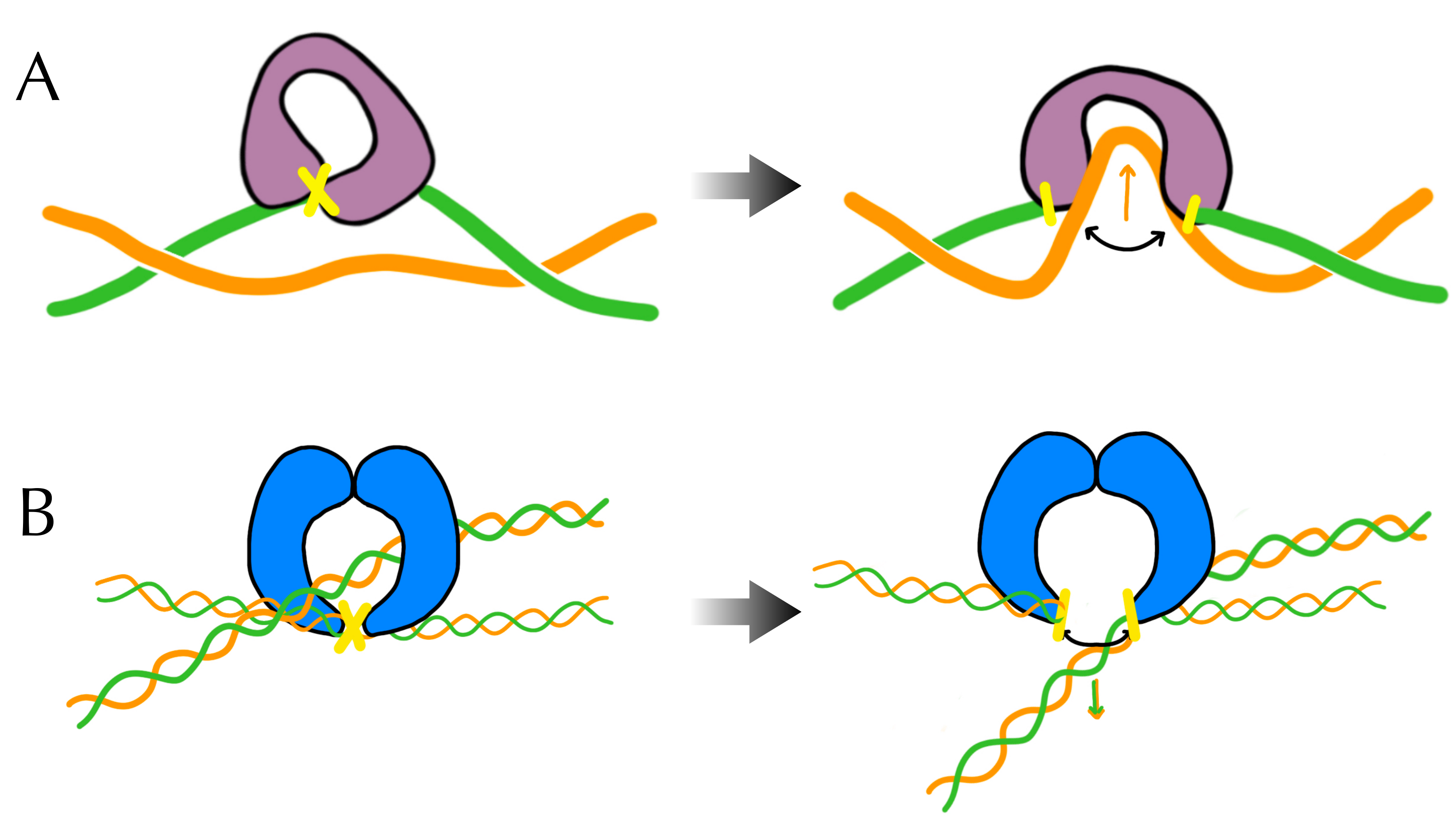}
\caption{{\corrprev Cartoon of the main step responsible for the change of the DNA's linking number during the enzymatic cycle of {\corr prokaryotic} Topo~I {\corr (TopA, equivalently)} and DNA gyrase -- {\lc DNA cuts are indicated in yellow.} A) {\corr TopA} scenario (adapted from~\cite{wang_cellular_2002,mckie_dna_2021}): the enzyme cuts a single strand (class~I topoisomerase) of DNA and makes the other strand pass through the cut before DNA is re-ligated ({\corr type IA}). {\corr Note that the prokaryotic Topo~I makes the twist change by a single unit, whereas the eukaryotic Topo~I makes the twist change by multiple units by allowing rotation of the uncut strand (type IB)}. B) DNA gyrase scenario (inspired from~\cite{nollmann_multiple_2007}): the enzyme cuts both strands of one DNA duplex (class~II topoisomerase) and makes {\lc another} duplex pass through the cut before DNA is re-ligated. The blue shapes {\lc indicates} a dimer of GyrB, {\lc omitting specific structural details}. The {\lc complete} complex involves an additional dimer of GyrA. {\lc  More detailed information on these structures and enzymatic cycles can be found in~\cite{mckie_dna_2021}.}}}
\label{fig:topo}
\end{figure}

{\corrprev
During transcription elongation {\it in vivo}, although direct evidence is currently lacking, numerous experiments suggest that an RNA polymerase (RNAP) generally undergoes minimal rotation around DNA (see section~\ref{sec:transcription}). Consequently, it generates both negative and positive supercoiling {\lc behind and ahead of it}. This supercoiling implies a restoring torque from DNA acting on the RNAP. Without the release of this torque, the RNAP would eventually stall~\cite{ma_interplay_2014} and transcription might terminate. This issue is resolved by topoisomerases. Specifically, evidence in various mesophilic bacteria points to a major role of Topo~I and DNA gyrase, which are able to respectively remove negative and positive supercoiling upstream and downstream the RNAP. The enzymatic reaction of {\corr the prokaryotic} Topo~I involves cutting one strand of the DNA duplex (class~I topoisomerase) and making the other strand pass through the cut (Fig.~\ref{fig:topo}A). This process introduces positive (+1) {\it twist} to the DNA molecule, which relieves the torsional stress associated with negative supercoiling. DNA gyrase, on the other hand, can adopt multiple modes of action~\cite{nollmann_multiple_2007}. In all cases, its enzymatic reaction involves cutting both strands of the DNA duplex (class~II topoisomerase) and making another duplex pass through the cut (Fig.~\ref{fig:topo}B). In ATP-consuming modes, via this process {\corr and an initial chiral wrapping of DNA~\cite{basu_dynamic_2018}}, DNA gyrase introduces negative (-2) {\it writhe} to the DNA molecule, which can then be rapidly converted into negative twist to alleviate the torsional stress associated with positive supercoiling. 

During DNA replication {\it in vivo}, numerous experiments also suggest that DNA polymerase rotates while advancing along the unreplicated DNA. This rotation nevertheless appears not be sufficiently rapid to prevent the accumulation of torsional stress in front of the replication machinery. Just as in transcription, the main topoisomerase that resolves this issue is DNA gyrase. Importantly, the rotation of DNA polymerase gives rise to an additional specific topological stress during the replication process, resulting in the newly synthesized DNA molecules twisting around each other and forming super-helices called precatenanes (section~\ref{sec:replication}). The resolution of these precatenanes is primarily carried out by a class~II topoisomerase known as Topo~IV, whose exact mechanism of action {\it in vivo} is still debated (section~\ref{sec:precat}).}

{\moveprev Altogether, while additional topoisomerases exist~\cite{wang_cellular_2002,forterre_origin_2007,mckie_dna_2021}, DNA gyrase, Topo~I, Topo~IV but also Topo~III (a class~I topoisomerase involved in decatenation of replicated DNA~\cite{mckie_dna_2021}) are considered as the most important topoisomerases in mesophilic bacteria. Notably, the average linking number of DNA in these bacteria has been shown to predominantly reflect the relative activity of Topo~I and DNA gyrase {\corrprev only}~\cite{drlica_control_1992,rovinskiy_rates_2012} {\corr as well as Topo~IV~\cite{zechiedrich_roles_2000}}.}

\iffalse

\begin{figure}[t]
\floatbox[{\capbeside\thisfloatsetup{capbesideposition={right,top},capbesidewidth=0.35\linewidth}}]{figure}[\FBwidth]
{\caption{{\corrprev Cartoon of the main step responsible for the change of the DNA's linking number during the enzymatic cycle of {\corr prokaryotic} Topo~I {\corr (TopA, equivalently)} and DNA gyrase. A) TopA scenario (adapted from~\cite{mckie_dna_2021}): the enzyme cuts a single strand of DNA and makes the other strand pass through the cut (class~I topoisomerase) before DNA is re-ligated. {\corr Note that the prokaryotic Topo~I makes the twist change by a single unit (type A), whereas the eukaryotic Topo~I makes the twist change by multiple units by allow rotation around the uncut strand (type B)}. B) DNA gyrase scenario (inspired from~\cite{nollmann_multiple_2007}): the enzyme cuts both strands of one DNA duplex and makes a distant duplex pass through the cut (class~II topoisomerase) before DNA is re-ligated. The blue shapes represent a dimer of GyrB. The full complex involves an additional dimer of GyrA.}}
\label{fig:topo}}
{\includegraphics[width=\linewidth]{./topoisomerase.pdf}}
\end{figure}

\fi

\subsection{{\corrprev Supercoiling density and its measurement} \label{sec:measure}}

{\corrprev What is the level of DNA supercoiling in bacteria? This simple question actually carries various subtleties related to measurement, particularly {\it in vivo} measurement. To comprehend this issue, let us briefly revisit the classical methodology used for supercoiling measurements in cells. First, it is important to note} that DNA supercoiling {\it is not measured on chromosomes, but on plasmids}. The latter are small circular DNA molecules of about 2-5 kilobase pairs (kb) that coexist with chromosomes and can be easily extracted from cells to quantify their {\corrprev linking number}. This quantification relies on the measurement of plasmid migration properties on gels as these are sensitive to the compaction status of plasmids and, hence, to their level of super-structuring. The tacit assumption that plasmids are good {\corrprev topological} proxies for chromosomes is then justified by the fact that topoisomerases are expected to behave similarly on both chromosomal and plasmid DNA.

Next, to compare DNA supercoiling levels between different bacteria, it is useful and customary to define the supercoiling density $\s$. This value is equal to the relative difference between the measured linking number and the linking number for the {\corrprev mechanically} relaxed state: $\s=\frac{\Lk-\Lk_0}{\Lk_0}$. The supercoiling density thus indicates the relative over- or under-winding of a DNA molecule with respect to the winding of a relaxed molecule (Fig.~\ref{fig:linking}). Namely, if $\Lk<\Lk_0$, the supercoiling density $\s$ is negative and the molecule has typically fewer helices than the corresponding relaxed B-DNA molecule, meaning that DNA is underwound with more base pairs per turn. Inversely, if $\Lk>\Lk_0$, $\s$ is positive and DNA is overwound, with less base pairs per turn. Let us recall, nevertheless, that part of the difference in local helicities between a supercoiled molecule and its relaxed counterpart takes the form of super-structuring (Fig.~\ref{fig:linking}) -- see section~\ref{sec:equilibrium} for further details.

Finally, in addition to being an indirect estimate of chromosomal supercoiling, reported values of supercoiling densities usually correspond to quantities that are averaged over a cell population. Assuming a homogeneous population, this is equivalent to averaging over time. In this context, the measured supercoiling densities have been found to be negative for mesophilic bacteria, with mean values not exceeding {\lc $-0.1$}~\cite{bliska_use_1987}. {\corr Note that it has been argued that this negative supercoiling is maintained by a proper balance of topoisomerase activity in the context of the regulation of gene expression~\cite{menzel_regulation_1983}, as negative values tend to favor transcription initiation (section~\ref{sec:transcription}).}

Altogether, these considerations mean that chromosomes of mesophilic bacteria are {\it predicted to be underwound on average, i.e., along the genome and over time}. More precisely, using the definition of $\s$, the number of base pairs per helix turn in the absence of writhe, denoted $n_\s$, verifies $\s=\frac{1/n_\s-1/n_0}{1/n_0}$, where $n_0\simeq 10.5$ is the corresponding number for torsionally relaxed B-DNA. Therefore, $n_\s \simeq 10.5/(1+\s)$ such that, for a typical measured value of $\s = -0.05$~\cite{bliska_use_1987}, $n_\s \simeq 11.1$ base pairs.

Let us nevertheless finish by noting that recent molecular techniques associated with DNA sequencing, such as Psora-seq~\cite{visser_psoralen_2022} or GapR-seq~\cite{guo_high-resolution_2021}, have paved the way for estimating supercoiling levels along chromosomes. {\corr Results show in particular that genomic distributions reflect transcriptional activities. Models aiming at predicting, or simply explaining these profiles, thus need to be developed in the context of transcription, in particular by including the specific action of topoisomerases (section~\ref{sec:transcription}).}

\section{Physical modeling of supercoiled DNA: fundamentals}
\label{sec:models}

{\corrprev
If bacterial genomes are relatively small compared to those of eukaryotes, chromosomes comprising several million base pairs are nevertheless gigantic macromolecules with contour lengths in the $\SI{}{mm}$ range, whose shapes undergo permanent changes due to thermal fluctuations and the action of the molecular machinery living organisms have evolved to structure, transcribe and replicate the genome. In this section, we introduce the notion of physical modeling and explain how such an approach applied at a resolution of the atoms comprising the DNA molecule, although feasible in principle, face unsurmountable difficulties on the possibility of brute-force modeling such gigantic macromolecules. We then explain how successive approximations, also known as coarse-grained descriptions, can be considered by dropping more and more details of the molecule. We introduce more particularly the rod-like chain model~\cite{vologodskii_conformational_1992,bouchiat_elastic_2000}, which is the simplest model for studying the folding properties of supercoiled DNA. For more details we refer the reader to the \hyperref[sec:app_models]{Appendix}, where we provide a more exhaustive introduction into the subject of physical modeling.}

\subsection{{\corrprev Atomistic modeling}}
%The paper being about physics based modelling we might as well commence by defining what we mean.

The prototypical example for physics based modeling is the work by Newton, who defined an equation of motion (the acceleration of a body is equal to the ratio of the force acting on it and its mass) and the ``force field'' describing the gravitational interaction between massive bodies like the sun, the earth and the proverbial apple. By solving these equations, Newton was able to explain that Kepler's laws of planetary motion in the sense that they emerge from this more fundamental description, which also describes the ballistic trajectory of a cannon ball on earth~\cite{weinberg2015explain}.

Conceptually, Molecular Dynamics Simulations~\cite{frenkel2001understanding,Karplus_McCammon_biomolecular_simulation_2002,brooks_CHARMM_2009,case2021amber} proceed on an atomic level along the same lines. It is often used in the framework of Statistical Mechanics (section~\ref{sec:statmec} of the Appendix) to explain or predict emergent macroscopic properties from the behavior of microscopic (e.g, atomic) states. The underlying equations of motions for the atoms are those of Newton, which are nowadays solved numerically for force fields modeling the bonded and non-bonded interactions between the atoms. The emergent properties for, say, a model of water are now phase diagrams or material constants like the viscosity describing the liquid phase. In principle, such molecular dynamics simulations are ideal tools for studying the complexities of biomolecular systems. With steady advances in available computer power and the performance of employed codes~\cite{shaw2014anton,abraham2015gromacs,eastman2017openmm,phillips2020scalableNAMD,plimpton_LAMMPS_2022}, they provide an ever more powerful ``computational microscope''~\cite{schulten_computational_microscope_2009,shaw_computational_microscope_2012} into biomolecular structures and processes. Of particular interest for this review is their ability to help rationalize the structural {\corr properties of supercoiled DNA molecules~\cite{mitchell_atomistic_supercoiled_DNA_2011}, that is, the different ways of distributing the linking number of a molecule {\lc between the twist and the writhe} (section~\ref{sec:basics}). In particular, both cryo-electron microscopy~\cite{Irobalieva_supercoiled_DNA_2015} and atomic force microscopy~\cite{pyne_base-pair_2021} have revealed a diversity of spatial conformations significantly larger than that initially thought, as well as a systematic presence of sharply bent DNA and kinks. Molecular dynamics simulations could confirm this diversity and further highlight the mechanisms associated with the local deformation of DNA~\cite{Irobalieva_supercoiled_DNA_2015,pyne_base-pair_2021}, such as the tendency of nucleobases located at sharp bends to adopt splayed configurations.}

Extending the domain of application of molecular dynamics simulations, which currently concern molecules of a few hundred base pairs, to the bacterial scale is however not feasible {\corr with current technology}. Namely, using a single GPU for a system composed of $10^6$ atoms, one can currently simulate on the order of 10 nanoseconds per day. While this allows reaching the microsecond scale in $100$ days, simulating an entire $5$ Mb long bacterial genome, which comprises on the order of $10^9$ atoms, over biologically relevant time scales remains elusive. For instance, simulating a 100-minute-long cell cycle would require the time elapsed since the extinction of the dinosaurs. Coarse-grained models~\cite{Saunders_Voth_coarse_graining_2013,dans_multiscale_2016,jewett_Moltemplate_2021}, which consist of dropping fine details below a given resolution to build simpler descriptions that capture properties above this resolution, are thus inevitably needed to rationalize and predict the structuring properties of DNA {\it in vivo}.

\subsection{\moveprev DNA fiber models}

\begin{figure}[t]
\centering
\includegraphics[width=0.75\linewidth]{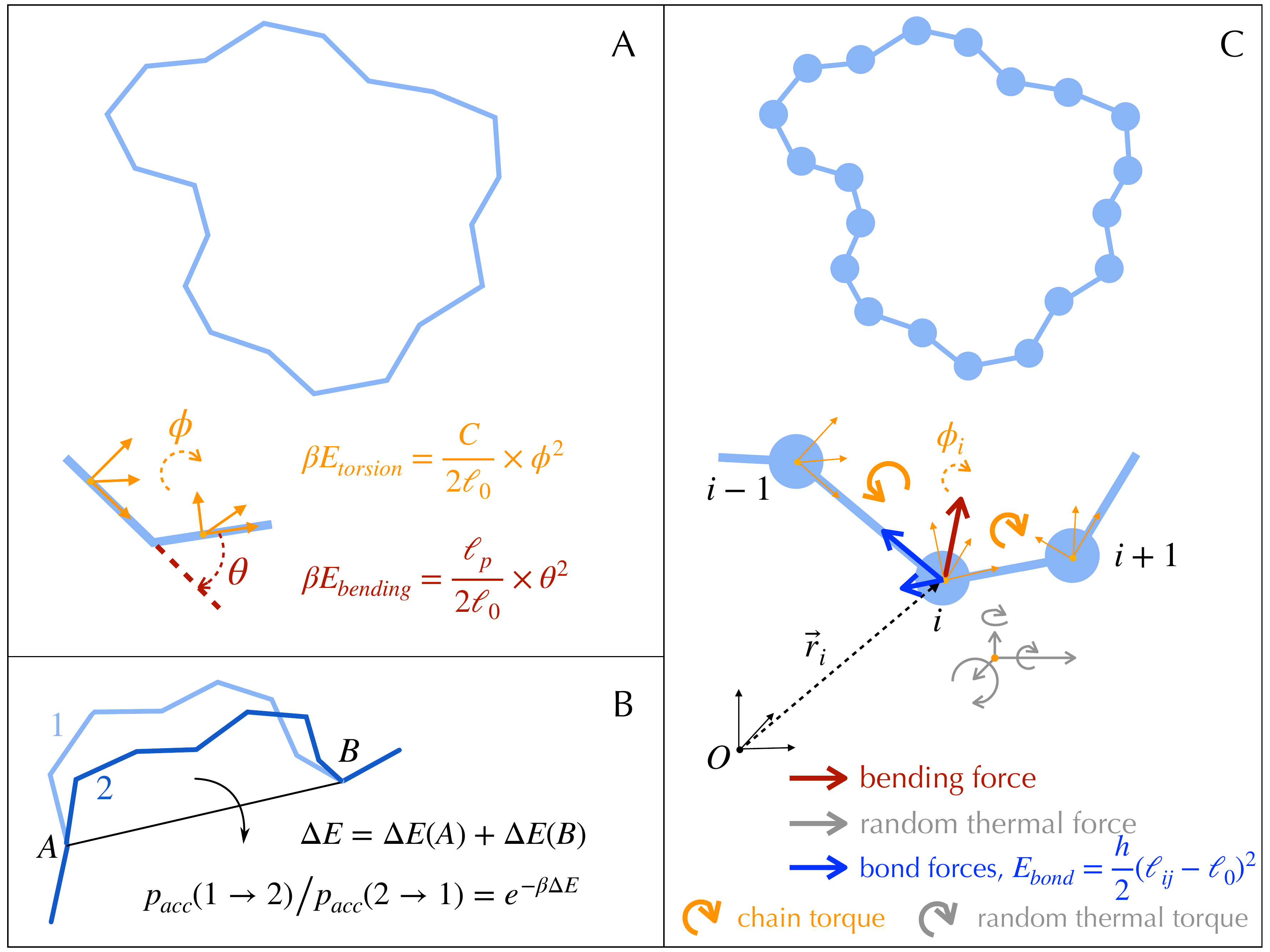}
\caption{{\corrprev Fiber models of DNA and numerical simulations. A) The simplest physical model of supercoiled DNA simplifies the DNA as a rod-like chain discretized into segments of a certain length $\ell_0$. The model involves only two independent parameters, the bending and the torsional moduli, which are respectively given by $\ell_p/\ell_0$ and $C/\ell_0$. $\ell_p$ is known as the bending persistence length and $C$ as the torsional persistence length. The associated energies are harmonic potentials of the bending angle $\theta$ and of the torsion angle $\phi$, respectively. The latter quantifies the rotational variation of the frames associated to each segment, {\lc a frame representing an orthonormal basis as indicated here by the orange vectors}. $\beta^{-1}=k_BT$ defines the thermal energy. B) A typical Monte Carlo simulation of the rod-like chain consists in iteratively rotating random groups of contiguous segments. The rotation of a specific group of segments is performed around the axes joining their flanking articulation points ($A$ and $B$). It is accepted depending on the associated variation of energy $\Delta E$, which involves only the articulation points, according to a probability rule (equation) that ensures reaching thermodynamic equilibrium at long time. C) DNA dynamics can be simulated using Brownian Dynamics methods. To this end, DNA is modeled as beads on a string. At each time step, the motion of each bead is updated according to its equation of motion, which involves frictional forces and torques from the solvent (not shown), forces and torques coming from the neighboring connected beads, and random forces and torques (thermal noise from the solvent).}}
\label{fig:models}
\end{figure}

{\moveprev Structural details of DNA can be coarse-grained to build single-nucleotide resolution polymer models~\cite{harris_modelling_2006,ouldridge_structural_2011,manghi_physics_2016}. Simulating such models is nevertheless still limited to less than $\SI{1}{kb}$ long molecules, calling for less resolved models to investigate structuring properties above the gene scale. In this regard, rigid base~\cite{lavery_ABC_2009,gonzalez2013sequence} or base-pair~\cite{olson1998_rigid_base_pair_model,becker2006indirect_readout,lavery_ABC_2009,becker2009dna_nanomechanics} models of B-DNA allow to preserve the sequence-dependent structure and elasticity of the canonical double-helix. Further coarse-graining~\cite{becker2007rigid,Maddocks_cgDNA_2014,sakaue_coarse_graining_DNA_2023} leads to tens-of-base-pairs resolution} {\corrprev fiber models of DNA (Fig.~\ref{fig:models}), which still preserve the microscopic mechanical properties of DNA. These fiber models have been used to address the properties of DNA molecules up to several tens kb (section~\ref{sec:3dmodel}). The rod-like chain model~\cite{vologodskii_conformational_1992,bouchiat_elastic_2000} is a prototypical example in which DNA is modeled as a series of articulated rigid segments (Fig.~\ref{fig:models}A). The relative orientation of consecutive segments is constrained by two parameters, the bending and torsional moduli, which quantify the resistance of DNA to bending and torsion, respectively. Importantly, fiber models neglect the specific structure of the double helix itself. As a consequence, they necessitate the inclusion of an effective treatment for conserving the linking number (section~\ref{sec:statmec} of the Appendix).}

\subsection{\corr Numerical simulations}
\begin{figure}[t]
\floatbox[{\capbeside\thisfloatsetup{capbesideposition={right,top},capbesidewidth=0.4\linewidth}}]{figure}[\FBwidth]
{\caption{{\corrprev Snapshot of a typical DNA conformation obtained through Monte Carlo simulation using the self-avoiding rod-like chain model, with a negative supercoiling density $\s=-0.06$. The DNA molecule considered here is $\SI{30}{kb}$ long, with each segment containing $\SI{30}{bp}$. The leftmost panel provides a zoomed-in view of a plectonemic super-structure, emphasizing the discrete nature of the segments.}}
\label{fig:plecto}}
{\includegraphics[width=\linewidth]{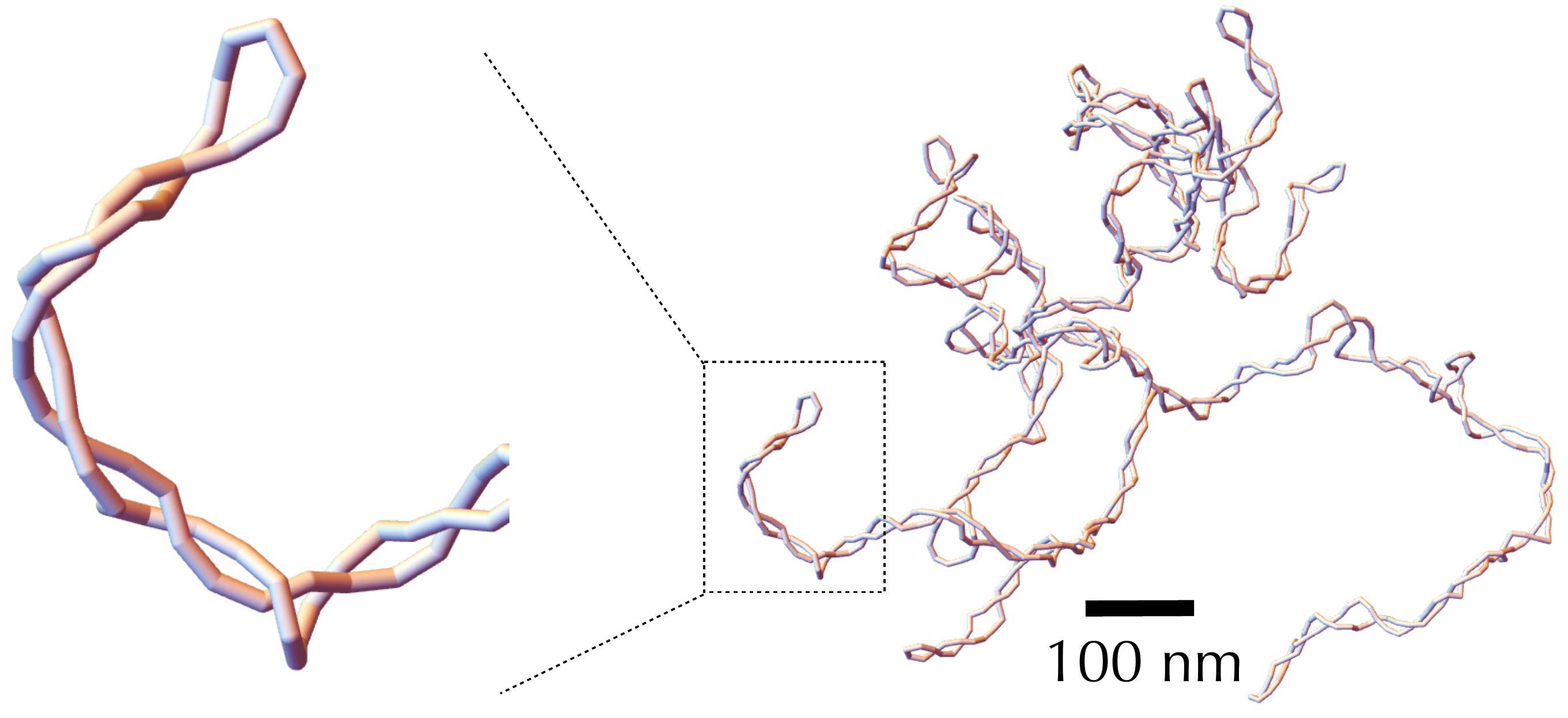}}
\end{figure}

{\corrprev
Solving analytically the simplest model as the rod-like chain model by predicting for instance the spatial extension of the molecule leads to unsurmountable difficulties. Anticipating phenomena where volume exclusion plays an important role such as in the presence of plectonemic DNA is also known to be a difficult task, although phenomenological approaches based on thermodynamic arguments have been proved to be particularly insightful (section~\ref{sec:eq_phen} of the Appendix). Numerical simulations of polymer chains are thus often necessary to investigate the folding properties of DNA fiber models. In this regard, equilibrium properties can be studied with the help of Monte Carlo simulations~\cite{frenkel2001understanding}, which allow to explore the space of possible DNA conformations often in an efficient way, with the help of a non-physical random dynamics (Fig.~\ref{fig:models}B). Dynamical properties can be studied using Brownian Dynamics simulations~\cite{allison_multistep_1984,chirico_calculating_1992}. To this end, the DNA chain is described in terms of beads~\cite{chirico_kinetics_1994,jian_combined_1997} (Fig.~\ref{fig:models}C) and its motion is simulated by considering the equations of movement for the beads. Namely, Brownian dynamics simulations assume that each DNA bead experiences a combination of friction and (correlated) random forces and torques coming from the solvent (cytoplasm) plus a combination of forces and torques coming from the translational and rotational motions of the connected neighbor beads along the chain (Fig.~\ref{fig:models}C). Applied to the rod-like chain model where the self-avoidance properties of the DNA chain is also considered, these simulations show that the three-dimensional folding induced by DNA supercoiling is indeed of a plectonemic type~\cite{vologodskii_conformational_1994} (Fig.~\ref{fig:plecto}). 
}

{\corr
\section{Structural predictions from equilibrium fiber-like models}
\label{sec:equilibrium}}

Molecular tools have been developed to probe the topology of DNA {\it in vivo} at multiple scales~\cite{lagomarsino_structure_2015}, with the recent possibility of obtaining information on the distribution of torsional stress along the genome~\cite{guo_high-resolution_2021,visser_psoralen_2022}. Yet, the {\it in vivo} occurrence and permanence of associated structural phenomena remain poorly quantified. Difficulties lie {\corrprev both in  the difficulty of measuring supercoiling densities (section~\ref{sec:measure}) and} in the small size of the structures involved (of the order of the nm). {\corrprev Many modeling questions have thus revolved around predicting the relative proportion of local deformation of the double helices and of super-structuring.} Equilibrium Statistical Mechanics {\corrprev (section~\ref{sec:eq_phen} of the Appendix)} has played a major role in this matter. {\corrprev In the following, we discuss in more details both outcomes of these approaches and} their relevance for {\it in vivo} situations, which is a consequence of the often ``near-equilibrium'' nature of phenomena. {\corrprev To this end, we first present one-dimensional models aiming at specifically capturing  the local deformations of the DNA duplex. We next present three-dimensional models aiming, {\it in fine}, at capturing both the local deformations of the DNA helix together with the overall folding of the molecule. Finally, we discuss how these models have recently been used to provide novel physical insights into the question of the nature of the topological barriers that have been detected {\it in vivo}.}

%Indeed, predictions from such equilibrium approach can be shown} to be relevant {\it in vivo} although processes such as transcription and replication operate {\corrprev far from} equilibrium, as they consume both energy (ATP) and mass (nucleotides). %From a modeling viewpoint, this raises the question of the relevance of equilibrium approaches to predict {\it in vivo} features associated with DNA supercoiling.
%This is a consequence of the often ``near-equilibrium'' nature of phenomena, as we explain below.

%Specifically,}  More, for a relaxed molecule, $\Tw_{relaxed}\simeq\Lk_0$ and $\Wr_{relaxed}\simeq 0$. Any torsional stress would then affect, from a thermodynamic standpoint, the most favorable base pairing geometry associated with B-DNA. As a result, there is a thermodynamic force in DNA that pushes towards conformations characterized by a twist $\Tw$ such that $\Delta \Tw=\Tw-\Lk_0=0$. However, in the presence of supercoiling ($\sigma \neq 0$), conservation of the linking number implies $\Wr=-\Delta \Tw + \s \Lk_0$. Therefore, $\Delta \Tw=0$ is associated with a large value of the writhe, $\Wr=\s \Lk_0$. The corresponding super-structure has a cost both in entropy, by reducing the number of possible conformations, and in bending energy. As a result, there is an opposite thermodynamic force that pushes towards conformations with low values of $\Wr$ or, equivalently, towards large values of $\Delta \Tw$.

\subsection{{\corrprev One-dimensional} models: {\corr predicting denaturation bubbles and other non-B DNA motifs}}
\label{sec:1dmodel}

{\moveprev The intensity of supercoiling-induced mechanical stress depends on the local DNA sequence. As a consequence, various phenomena can take place at specific locations along the genome. These include DNA denaturation as shown by the pioneering work of Vinograd and his collaborators in the 1960s~\cite{vinograd_early_1968}, generation of DNA forms alternative to B-DNA~\cite{mirkin_discovery_2008} and generation of alternative secondary DNA structures such as cruciforms~\cite{mizuuchi_cruciform_1982}. Importantly, some of these structural motifs have a functional role, making the {\it physical prediction} of their occurrence and distribution along the genome an important {\it biological problem}~\cite{du_genome-wide_2013}.}

\begin{figure}[t]
\centering
\includegraphics[width=0.75\linewidth]{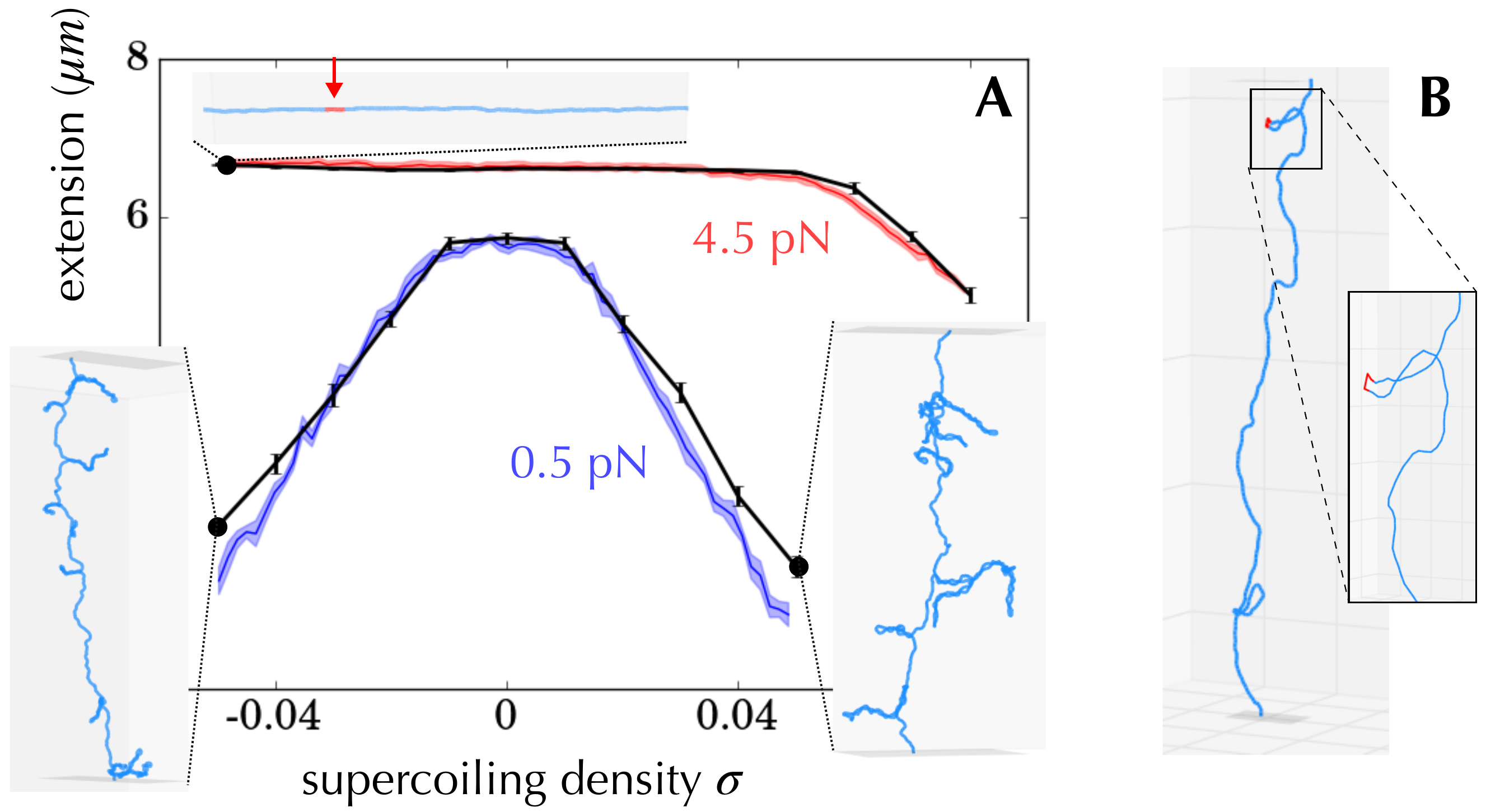}
\caption{Example of diversity of structures obtained using a coarse-grained model of DNA at a resolution of $\SI{10}{bp}$, including the possibility of forming alternative DNA structures such as denaturation bubbles~\cite{lepage_polymer_2019}. A) Comparison between experiments~\cite{vlijm_experimental_2015} (colored curves) and simulations (black curves) for a $\SI{21}{kb}$ long molecule manipulated by a magnetic tweezer. The $x$-axis indicates the imposed supercoiling density on the molecule, and the $y$-axis shows the measured extension of the molecule. The experiments were conducted at two forces ($\SI{0.5}{pN}$ in blue and $\SI{4.5}{pN}$ in red). The inner panels show the typical conformations of the molecule obtained in the simulations for different experimental parameters (black dots). For example, the top panel indicates that when the molecule is stretched at $\SI{4.5}{pN}$ and undergoes a negative supercoiling of $\sim -0.04$, a denaturation bubble forms (indicated in red). The other conformations indicate the presence of plectonemes. B) For some force and supercoiling density, conformations can display denaturation bubbles (in red) located at the apex of plectonemes. These were initially predicted to occur using a coarse-grained polymer model of DNA at the nucleotide level~\cite{matek_plectoneme_2015}.}
\label{fig:thermo}
\end{figure}

%An illustrative example of the power of equilibrium statistical mechanics methods concerns the question of the alternative forms of DNA that are generated during the deformation of the double helix. 
How are Statistical Mechanics models built to {\corr address the problem of the tendency of a given subsequence of DNA to denature or form alternative forms?} First, they most often 
%The most common approaches are based on two assumptions. 
neglect the effects of writhe, which is similar to assume a stretching force of a few pN (Fig.~\ref{fig:thermo}), so that the problem becomes one-dimensional~\cite{fye_exact_1999}. In doing so, analytical calculations are possible, making it possible to establish mathematical relationships between observables (i.e., measurements performed on the system) and system parameters (e.g., supercoiling level). It is then possible to predict behaviors without resorting to simulations which are often time-consuming and limited from the viewpoint of exhaustivity. Second, {\corrprev most approaches assume} that supercoiling constraints are relaxed much faster than they are produced ({\it near-equilibrium condition}). This hypothesis {\corrprev is justified}, for example, in the case of transcription, whose initiation step requires the formation of a DNA denaturation bubble~\cite{murakami_bacterial_2003,mejia-almonte_redefining_2020}. Namely, the twist and writhe relaxation times (below $\SI{1}{ms}$) for a 10 kb long molecule are typically {\corrprev four} orders of magnitude smaller than the time for synthesizing a 1 kb long messenger RNA ({\corrprev $\geq\SI{10}{s}$})~\cite{joyeux_requirements_2020,fosado_nonequilibrium_2021,wan_two-phase_2022} and {\corrprev one} order of magnitude with respect to the time for synthesizing a single base pair. Thus, questions concerning the energy required to denature DNA have been systematically addressed in the context of the equilibrium statistical mechanics of one-dimensional systems ~\cite{benham_torsional_1979,fye_exact_1999,jost_bubble_2011}. In particular, efficient semi-analytical approaches allow to predict the most probable sites of denaturation at the scale of a genome~\cite{jost_bubble_2011,jost_twist-dna_2013}.

Despite simple assumptions with respect to the complexity of {\it in vivo} phenomena, including the neglect of super-structuring, {\corrprev these equilibrium one-dimensional approaches have been shown to be} sufficiently predictive to be used, for example, in the analysis of the sensitivity to supercoiling of transcription initiation~\cite{forquet_role_2021}, {\corr in accord with the necessity of DNA to denature at the promoter (see section~\ref{sec:transcription} for insights)}. This suggests that strong deformations of the double helix is often dominant {\it in vivo} {\corrprev and, hence, that forces} on the pN range are expected to act on bacterial DNA~\cite{strick_stretching_2003}. {\corr Along the same line, these approaches have been used to predict the appearance and location of non-B DNA motifs~\cite{wang_stress-induced_2004,du_genome-wide_2013}, which appears to reflect the role of supercoiling in the regulation of transcription~\cite{du_genome-wide_2013}.}

\subsection{Three-dimensional structuring: from the first observations to the first polymer models}
\label{sec:3dmodel}

{\corrprev In addition to demonstrate that supercoiled DNA denatures, Vinograd and his collaborators used electron microscopy experiments to reveal, for the first time, the capacity of (viral) circular DNA molecules to form super-structures~\cite{vinograd_twisted_1965,vinograd_physical_1966}. Notably, they observed ``plectonemes'', while the term would be only coined in the late 1980s (see~\cite{vologodskii_conformational_1992}). Remarkably, they attributed these super-structuring properties to invariant topological properties of circular molecules: ``In closed double circular DNA, the number of degrees of angular rotation of one strand around the other is invariant''~\cite{vinograd_physical_1966}. Several years later,} 
%which they rationalized as 
%Soon after the findinds that supercoiled DNA 
%Another illustrative example of the power of equilibrium statistical mechanics approaches concerns the problem of the three-dimensional structure adopted {\it in vivo} by bacterial DNA. 
in the 1970s, pioneering electron microscopy experiments {\corrprev revealed that the bacterial chromosome extracted from \Ecoli\ cells} was also made of plectonemes~\cite{delius_electron_1974,kavenoff_electron_1976} but also of numerous loops~\cite{delius_electron_1974}. In 1990, {\it in vitro} experiments, still visualized by electron microscopy, showed for {\it in vivo} relevant values of supercoiling density a systematic tendency of bacterial DNA to form plectonemes at small scales and trees at large scales~\cite{boles_structure_1990}.

{\corrprev These results raised fundamental questions, starting with the physical mechanisms behind the formation of plectonemic structures. In particular, since} an excess of writhe could manifest itself in the form of solenoids, {\corrprev how to explain the prevalence of plectonemes?} This question remained unanswered for many years before being partially solved in the early 1990s with the help of the first polymer models of {\corrprev supercoiled DNA at a resolution of a few tens base pairs}~\cite{klenin_computer_1991,vologodskii_conformational_1992}. These models, which are still at the basis of current works, account for the electrostatic repulsion of DNA (self-avoidance), the energies of DNA {\corrprev bending} and torsion, which result from a coarse-grained description of DNA that neglects fine atomic details, as well as the global constraint of the conservation of the linking number (section~\ref{sec:models}).
% -- the latter can be implemented locally~\cite{carrivain_silico_2014,lepage_modeling_2017}, i.e., efficiently through the notion of parallel transport~\cite{bergou_discrete_2008}.
In 1994, the question was definitively resolved by Marko and Siggia on the basis of a quasi-analytical {\corrprev solution of a phenomenological equilibrium thermodynamics description} of these {\corrprev microscopic} models~\cite{marko_fluctuations_1994}, showing that under physiological conditions of temperature, salt and supercoiling density, plectonemes are thermodynamically favored compared to solenoids. The reason lies in  the ``large'' energy of {\corrprev bending} of solenoids, which can be reduced drastically in plectonemes while keeping similar torsional stresses~\cite{marko_fluctuations_1994}. Single-molecule magnetic tweezers experiments combined with fluorescent labelling of DNA~\cite{van_loenhout_dynamics_2012} and polymer simulations~\cite{lepage_thermodynamics_2015} have then shown that the length of plectonemes {\it in vitro} are on the order of $\SI{1}{kb}$. Electron microscopy~\cite{boles_structure_1990} and statistical mechanics of plectonemes~\cite{marko_statistical_1995,barde_energy_2018} also revealed a diameter of the plectoneme varying between $\simeq\SI{30}{nm}$ at $\s\simeq-0.025$ and $\simeq\SI{5}{nm}$ at $\s\simeq-0.1$.
% — the same work also provided a rationale for reported ratio of 3 to 1 for the excess of writhe with respect to the excess of twist~\cite{boles_structure_1990}.
{\corrprev Finally,} due to the entropic contribution of branches, {\corrprev Marko and Siggia} further showed that plectonemic structures become branched and form trees at large scales~\cite{marko_statistical_1995}, {\corrprev rationalizing} both experiments~\cite{boles_structure_1990} and numerical simulations~\cite{vologodskii_conformational_1992}. 

\subsection{\corrprev A simulation toolbox to anticipate structuring properties of bacterial DNA.} 

The pioneering work of Vologodskii and his collaborators in the early 1990s, which focused on the development of Monte Carlo simulations {\corrprev (section~\ref{sec:simus})} for topologically constrained polymer chains~\cite{vologodskii_conformational_1994}, sparked intense and ongoing research on the thermodynamic properties of supercoiled DNA at the scale of a few kb, {\corrprev typically up to a few tens kb} (see Fig.~\ref{fig:thermo} for an example). The self-avoiding rod-like chain model~\cite{vologodskii_conformational_1992} {\corrprev (Fig.~\ref{fig:models})}, also known as the twistable worm-like chain model~\cite{nomidis_twist-bend_2019}, is typical of this approach and has been instrumental in analyzing the {\corrprev equilibrium} folding properties of both positively and negatively supercoiled DNA molecules without strong deformation of the B-DNA double helix. These properties include molecular extensions~\cite{vologodskii_extension_1997}, torques~\cite{schopflin_probing_2012,lepage_thermodynamics_2015}, and conformation details of super-structures~\cite{vologodskii_conformational_1992,bednar_twist_1994,klenin_modulation_1995}.
%, as obtained in magnetic tweezer experiments and cryo-electron microscopy. 
The models can be extended to include DNA denaturation and the formation of alternative forms that occur at high negative supercoiling levels~\cite{lepage_polymer_2019}. Brownian dynamic simulations of supercoiled DNA {\corrprev (Fig.~\ref{fig:models}, section~\ref{sec:simus})} were also developed in the early 1990s by Langowski and his collaborators, enabling the study of the dynamical properties of DNA loci~\cite{chirico_kinetics_1994}. 

{\corrprev This toolbox of polymer simulations has been used for more than 30 years not only to rationalize experimental but also to anticipate possible non-trivial properties of supercoiled DNA. An illustrative example comes from an early study by Langowski and collaborators}. Namely, 
%Many works have been performed to capture various structuring multiple aspects of supercoiled DNA Instead of trying to cover It is not possible to cover all the research that has been conducted in this field within the scope of this review. Therefore, we will only highlight a few results that we consider particularly relevant for understanding DNA structuring {\it in vivo}. For instance, 
their Brownian dynamics simulations predicted in the late 1990s that plectonemes should move along the DNA not only through (slow) diffusion but also by disappearing at one location to reappear at a distant location {\corr along the DNA}~\cite{chirico_brownian_1996}. This ``hopping'' type of motion was observed years later in fluorescent-labelling single-molecule experiments {\corr for supercoiled DNA stretched by pN range forces~\cite{van_loenhout_dynamics_2012} and, hence, is expected to occur as well {\it in vivo} -- plectoneme hopping in~\cite{van_loenhout_dynamics_2012} could be distinguished by the concomitant disappearance of a fluorescence spot (associated with the high density of a plectoneme) and appearance of another spot along the molecule}. Brownian dynamics simulations {\corrprev further revealed}, for molecules of a few kb, that loci tend to make contacts through intra-plectoneme slithering (secondary type of contacts) rather than through inter-plectoneme random collisions (tertiary type of contacts)~\cite{huang_dynamics_2001} -- tendency that may be reinforced by the hopping motion of plectonemes. The genomic range for which secondary contacts are expected to be more frequent {\corrprev than tertiary contacts} {\it in vivo} nevertheless remains open. From a modeling viewpoint, this would require in particular to properly investigate finite size effects knowing that the size of molecules {\corrprev in simulations} are at most on the order of a few tens kb, i.e., {\corrprev two} orders of magnitude smaller than e.g.~the chromosome of \ecoli.

\subsection{\corrprev Topological barriers: insights from Physics} 

\begin{figure}[t]
\centering
\includegraphics[width=0.75\linewidth]{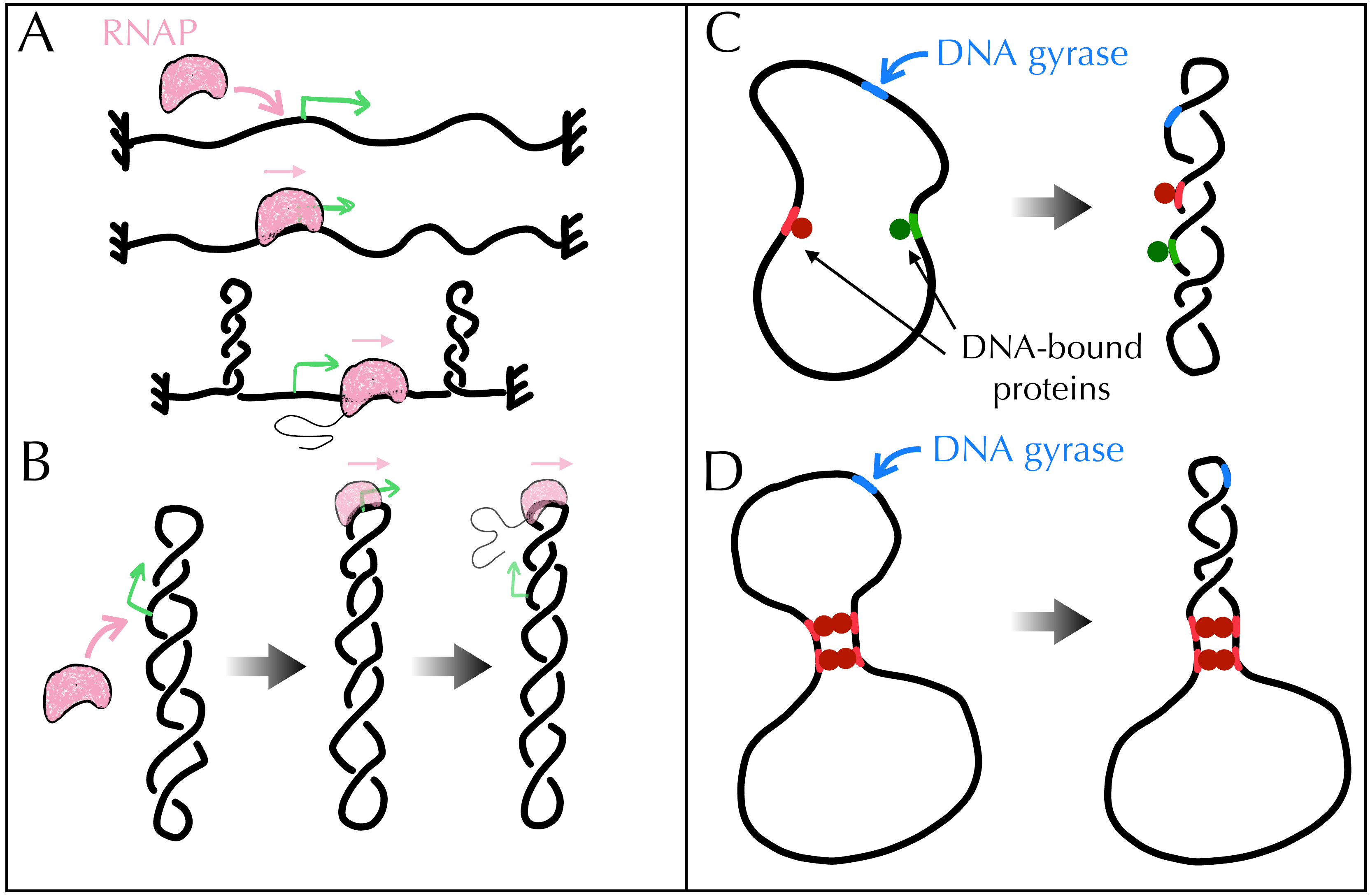}
\caption{{\corrprev Models of topological barriers as suggested by polymer simulations of supercoiled DNA in the presence of various proteins and enzymes. A) Simulations of an active process where an RNAP generates supercoils in a topologically constrained domain~\cite{fosado_nonequilibrium_2021} suggest that the RNAP itself can function as a topological barrier. In~\cite{fosado_nonequilibrium_2021}, the generated supercoils would indeed relax under the form of plectonemes that occur ``far'' from the RNAP. The RNAP thus prevents the mixing of the topological properties of the upstream and downstream DNA regions. B) Other studies~\cite{racko_transcription-induced_2018,joyeux_models_2022} have reported a tendency for a translocating RNAP to localize at the apex of plectonemes. In this case, the RNAP does not act as a topological barrier since the upstream and downstream DNA segments are intermingled. C-D) To evaluate the potential of DNA-bound proteins (green and red disks) to act as topological barriers, a possible experimental setup consists in considering a plasmid with a strong gyrase binding site (in blue) and in checking {\lc whether} gyrase activity at this site causes the entire plasmid, or only the region flanked by the binding sites of the proteins, to adopt a plectonemic super-structure~\cite{leng_dividing_2011}. C) Numerical simulations show that proteins that do not bridge DNA, even if they impede twist diffusion, are incapable of acting as topological barriers~\cite{joyeux_requirements_2020}. D) Proteins that block the diffusion of both twist and writhe, as in the presence of multiple successive bridges~\cite{leng_dividing_2011}, effectively operate as topological barriers~\cite{joyeux_requirements_2020}.}}
\label{fig:barriers}
\end{figure}

{\corrprev Contrary to a naive vision of a topological constraint (the linking number) acting on the chromosome as a whole, molecular genetics experiments, genetic recombination assays and electron microscopy of isolated chromosomes have revealed} that the genomes of 
\ecoli\ and {\it Salmonella} are actually organized into topologically independent domains whose size is on the order of $\SI{10}{kb}$~\cite{postow_topological_2004,deng_organization_2005}. Comparative genomics {\corrprev further} predicts this organization to be ubiquitous in bacteria and to be associated with the basal coordination of transcription~\cite{junier_conserved_2016,junier_universal_2018}. Yet, the nature of the topological barriers {\corrprev associated with this partitioning have remained highly debated. In this regard, polymer simulations have provided  insights into the possible implication of several factors. 

First, experimental results obtained {\it in vivo} from genetic recombination assays sensitive to the formation of plectonemes~\cite{higgins_surveying_1996} strongly suggest} that transcribing RNAPs behave as such topological barriers~\cite{deng_transcription-induced_2004,higgins_rna_2014}. A possible rationale {\corrprev could come from} active processes, {\corrprev i.e., from situations far from equilibrium (section~\ref{sec:noneq}) of the Appendix}. Namely, recent Brownian dynamics simulations have shown that the DNA supercoiling introduced by a transcribing RNAP might relax under the form of plectonemes that form far from the RNAP~\cite{fosado_nonequilibrium_2021} (Fig.~\ref{fig:barriers}A). The absence of any plectoneme {\corrprev embedding} the RNAP is indeed in accord with the capacity of this RNAP to block the diffusion of writhe. {\corrprev Yet,} other modeling works have reported the tendency for a transcribing RNAP to locate at the apex of plectonemes~\cite{racko_transcription-induced_2018,joyeux_models_2022} (Fig.~\ref{fig:barriers}B), which is in opposition with its functioning as a topological barrier. {\corrprev The fundamental reason for the difference between the outcomes of these far from equilibrium models remain to be elucidated}. Nevertheless, it is worth noting that the apical localization of plectonemes is consistent with previous {\it in vitro} experimental results~\cite{ten_heggeler-bordier_apical_1992}. {\corrprev Moreover, experimental studies indicate that a transcribing RNAP enhances the flexibility of DNA (see references in~\cite{ten_heggeler-bordier_apical_1992}), while modeling studies have shown that the most flexible part of DNA tends to preferentially localize at the apex of plectonemes~\cite{matek_plectoneme_2015,lepage_polymer_2019} (Fig.~\ref{fig:thermo}B).}

{\corrprev Second, {\it in vitro} experiments combined with genomic analyses of protein binding sites suggest the participation of certain nucleoid associated proteins such as H-NS~\cite{hardy_genetic_2005,noom_h-ns_2007}. H-NS is indeed able to bridge DNA to form loops~\cite{dillon_bacterial_2010} and, as demonstrated in the case of LacI, GalR or $\lambda$ O~\cite{leng_dividing_2011}, these loops may define topological domains. The problem, then, is to identify the conditions DNA-bridging proteins} must follow to be able to topologically insulate a genomic domain from its neighbor. In this regard, recent Brownian dynamics simulations have shown that not only bridges must block the diffusion of twist but they must also prevent DNA segments to rotate with respect to each other, i.e., they must block the diffusion of writhe, too~\cite{joyeux_requirements_2020} {\corrprev (Fig.~\ref{fig:barriers}C-D)}. %{\corrprev Looping in the case of H-NS might actually be transitory. This would be in accord with the reported stochastic nature of barriers~\cite{postow_topological_2004}. This would also explain the absence of looping signature in chromosome conformation capture data (see e.g.~\cite{lioy_multiscale_2018}).}

{\corrprev How, then, to systematically test the ability of DNA-bridging proteins to create topological barriers? Experimental insights for the transcription factors LacI, GalR or $\lambda$ O have already been provided. The method consisted in combining biochemical techniques and atomic force microscopy to study folding properties of plasmids both in the presence of multiple binding sites of such proteins and under the action of DNA-nicking and gyrase activities~\cite{leng_dividing_2011}. An interesting alternative approach could consist in exploiting fluctuation properties of supercoiled molecules. {\corr  Specifically, the variance of the extension of a molecule, as a function of both its supercoiling density and the intensity of a stretching force acting on it, can be accurately predicted using a phenomenological approach~\cite{skoruppa_equilibrium_2022}. Next, while the average extension can be shown to be insensitive to the presence of a bridge within the plectonemes, the variance depends on the location of the bridge~\cite{vanderlinden_dna_2022}. In a proof of concept study, this property has been utilized to experimentally identify the position of topological barriers created by two-site-specific} DNA restriction enzymes (whose cleavage was impeded). This was achieved by combining single-molecule experiments, Monte Carlo polymer simulations of supercoiled DNA, and analytical approaches~\cite{vanderlinden_dna_2022}.

%This innovative type of analysis is expected to not only aid in the investigation of protein binding to DNA but also provide critical new insights into the molecular actors involved in the topological organization of bacterial genomes.

%In summary, thermodynamic equilibrium approaches are a useful tool to study the structuring properties of DNA within a well-established framework. {\corrprev Predictions are expected to be relevant for many {\it in vivo} situations as a consequence of the often ``near-equilibrium'' nature of phenomena, mainly reflecting the high relaxation speed of topological constraints compared to the rate at which they are produced.} Using this approach, properties and sequence effects that occur at multiple scales can be studied as a function of parameters such as the supercoiling density or mechanical forces acting on DNA (Fig.~\ref{fig:thermo}).

\subsection{\corrprev The need for further coarse-graining the DNA fiber models} 

 {\corrprev At a larger scale, can fiber models of DNA} {\moveprev be used to simulate the folding of an entire bacterial chromosome?
Supposing that thermodynamic equilibrium is relevant for the large scale organization of chromosomes, which should be the case for sufficiently slow cell growth, the question at hand is how long a simulation must run to reach thermodynamic equilibrium. To that end, we can consider the most effective Monte Carlo methods for forming and equilibrating supercoiled DNA structures, which involve chain deformations that are particularly well-suited to relaxing plectonemic structures~\cite{liu_efficient_2008}. Simulations suggest that the characteristic number of iterations required to reach equilibrium in this context is of the order of the chain length ($L$)~\cite{liu_efficient_2008}. Suppose, then, that the topological constraint of the conservation of the linking number is implemented locally\cite{carrivain_silico_2014,lepage_modeling_2017},
% through the notion of parallel transport~\cite{bergou_discrete_2008}, which avoids to calculate the writhe of the molecule (which scales as $L^2$). In this context,
simulations show that $K$ elementary Monte Carlo moves (whose subchain sizes range from 1 to $L$) take a CPU time that scales as $K\times L^{1.2}$~\cite{lepage_thermodynamics_2015}. Assuming that this time can be reduced to $K\times L$ (the exponent $1.2$ reflects the management of the self-avoiding constraint), since $L$ moves are necessary to reach equilibrium, the characteristic simulation time for the most efficient simulations should scale as $L^2$ -- note that these simulations are challenging to parallelize due to non-trivial self-avoidance constraints~\cite{krajina_large-scale_2016}.

Knowing that it takes about $5$ hours on a 3.5~Ghz processor to reach equilibrium for a chain of $\SI{20}{kb}$ when the supercoiling density is not too intense (e.g., for $\sigma=-0.03$)~\cite{lepage_thermodynamics_2015}, the time to reach equilibrium for a $\sim\SI{500}{kb}$ long genome, such as the JCVI-syn3A synthetic minimal genome for which Hi-C data is available\cite{gilbert_generating_2021}, is of the order of $5\times(500/20)^2\simeq3000$ hours, or approximately $130$ days. For \ecoli, the time is approximately $35$ years. To scale up to chromosomes, particularly those with a length of a few Mb as that of \ecoli, coarser-graining methods that neglect the details of plectonemes are thus necessary. {\corrprev In section~\ref{sec:scaleup},} we discuss two main types of models resulting from these procedures: trees and bottle brushes.
}

\section{Supercoiling constraints and transcription}
\label{sec:transcription}

Awareness of the central role of DNA supercoiling in transcription dates from the 1970s~\cite{wang_interactions_1974}, with the seminal work of James C. Wang, who discovered the first topoisomerase~\cite{wang_interaction_1971} known today as Topo~I {\corrprev -- DNA gyrase was discovered five years later~\cite{gellert_dna_1976}}. In particular, Liu and Wang hypothesized that the most frequent situation in bacteria for a transcribing RNAP is to generate supercoiling stresses on each side of it because of the impossibility of the RNAP to rotate around DNA~\cite{liu_supercoiling_1987}. More precisely, because the transcribing RNAP and its associated mRNA interact with other macromolecules (ribosomes, regulatory factors, other RNAPs or DNA itself through e.g.~the formation of R-loops~\cite{thomas_hybridization_1976}), the resulting macro-complex experiences torsional friction. This hinders the rotation of the RNAP around DNA. In addition, DNA itself interacts with various macromolecules (e.g. membrane~\cite{lynch_anchoring_1993}, clusters of RNAPs~\cite{stracy_live-cell_2015}), which is expected to hinder its global rotation, too. As a consequence of the difficulty of both RNAPs and DNA molecules to rotate and according to the topological considerations of Fig.~\ref{fig:twin_scheme}, Liu and Wang surmised that the transcription of genes most often generates negative {\corrprev and positive} DNA supercoiling {\corrprev upstream and, respectively, downstream} the transcribing RNAPs, which they demonstrated for a particular case on a plasmid~\cite{wu_transcription_1988}. The corresponding {\it biological} model is known as the twin transcriptional-loop (TTL) model~\cite{liu_supercoiling_1987}. {\corrprev It is nowadays at the foundation of all {\it physical} models of the interplay between transcription and DNA supercoiling (Fig.~\ref{fig:trans_model}). In the following, we thus explain the ingredients and outcomes of these models, and discuss the open problems to be solved.}

\subsection{{\corrprev On including topoisomerase activity in physical models: some numbers}}
\label{sec:inctopo}

{\corrprev First, can a model based on the interplay between only DNA and RNAPs capture quantitatively gene transcription? To answer this question, let us recall a few numbers associated with gene transcription 
%To get a quantitative apprehension of the constraints at play during the transcription of a gene, let us recall that 
in the most-studied bacterium.} In \ecoli, RNAPs transcribe at a rate between $25$ and $\SI{100}{bp}$.s$^{-1}$, depending on the growth rate of the bacterium~\cite{bremer_modulation_2008}. In the extreme case of an absence of rotation of the RNAP, this means that the DNA unwinding associated with transcription generates between $\sim 2$ to $\sim 10$ positive (negative) supercoils per second upstream (downstream) the RNAP --  considering one supercoil per transcription of $\sim 10$ base pairs or one turn of the DNA double helix.
Considering the presence of topological barriers located at a distance of $\sim\SI{10}{kb}$ ($\sim\SI{1000}{supercoils}$) that prevent the dissipation of these supercoils~\cite{postow_topological_2004}, according to the TTL model, transcription activity is expected to make DNA supercoiling density $\s$ vary by an amount of at least $0.01$ every second on each side of the transcription complex. With respect to DNA, the effects of supercoiling become significant for $|\s|=0.01$ and highly disruptive for $|\s|=0.1$~\cite{strick_stretching_2003}. With respect to RNAPs, single-molecule studies have suggested that they stall {\it in vivo} for torques ($\Gamma$) on the order of $\SI{18}{pN}$~\cite{ma_transcription_2019} or equivalently, $|\sigma| \simeq 0.06$ -- using $\sigma = \Gamma/A$ where $A= \SI{300}{pN}$ is an average of the values estimated from single-molecule experiments for the regime where plectonemes are present ($\SI{200}{pN}$) and for the regime where super-structuring is absent ($\SI{400}{pN}$)~\cite{marko_torque_2007}. RNAP translocation along the DNA can then resume only if the associated torques are released, which can occur {\it in vivo} through two mechanisms: i) another RNAP compensates the supercoiling, which however does not solve the problem upstream and downstream the train of RNAPs; ii) topoisomerases relax supercoiling. In other frequent situations such as those involving divergent genes, supercoiling densities may actually vary even more abruptly. Namely, for two divergent promoters separated by a distance of $\simeq \SI{200}{bp}$, the transcription of the upstream gene would create a transitory barrier and the total variations of supercoiling would be on the order of $0.1$ every second.

{\corrprev Altogether, these numbers show that topoisomerase activity is required for transcription to properly proceed as soon as the elongating complex slowly rotates around DNA. Actually,} an often overlooked ingredient of the TTL model is the {\corrprev inclusion} of topoisomerases. Liu, Wang and collaborators indeed demonstrated that Topo~I and DNA gyrase were responsible for relaxing the upstream negative supercoils and the downstream positive supercoils, respectively~\cite{wu_transcription_1988}. They then anticipated that for gene expression to be properly predicted, one would need to include the activity of these topoisomerases~\cite{liu_supercoiling_1987,wu_transcription_1988}. {\corr 30} years later, {\corrprev not only} experiments have convincingly demonstrated that gene context {\corr plays a role in gene expression} as important as transcription factors~\cite{yeung_biophysical_2017,nagy-staron_local_2021,scholz_genetic_2022}, {\corr but they have} {\corrprev also} corroborated the relevance of the TTL model {\corrprev and} the {\it necessity} to consider Topo~I and DNA gyrase to quantitatively apprehend transcription~\cite{chong_mechanism_2014, ahmed_transcription_2017, yeung_biophysical_2017, kim_long-distance_2019, rani_genome-wide_2019, ferrandiz_genome-wide_2021, sutormin_interaction_2022, boulas_assessing_2023}.

\begin{figure}[t]
\centering
\includegraphics[width=0.75\linewidth]{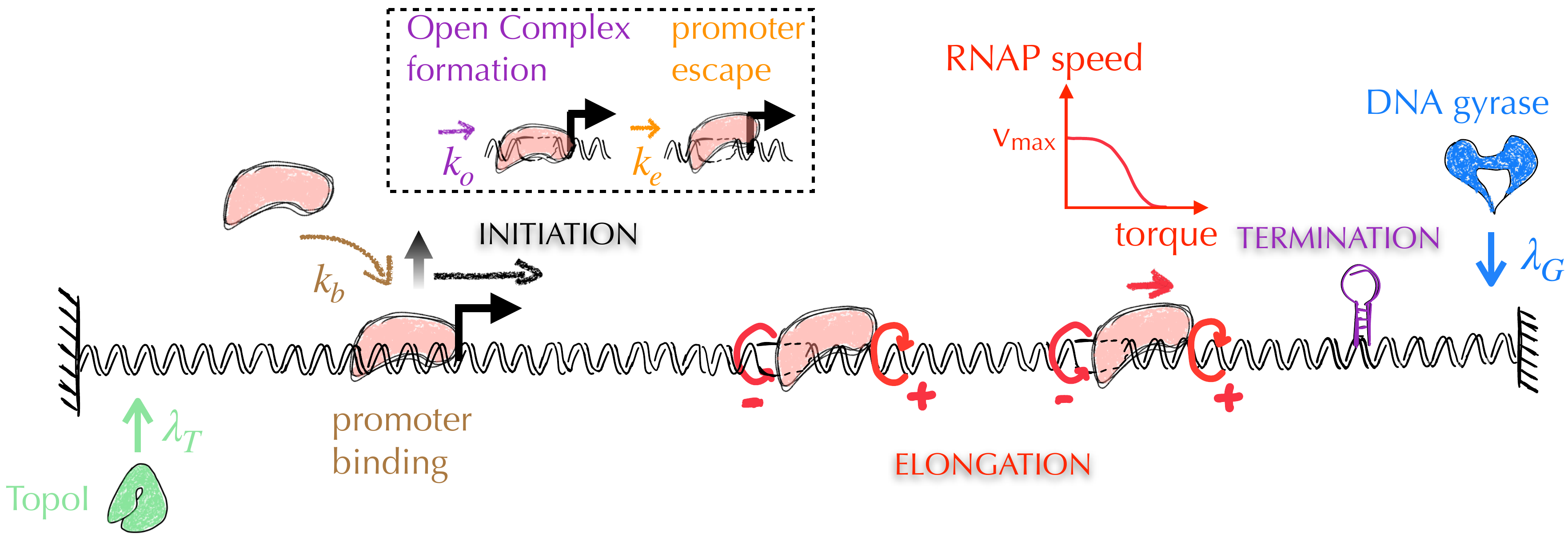}
\caption{{\corrprev Typical architecture of physical models of gene transcription {\corr -- for simplicity, we omit the representation of mRNAs}. The most recent models of gene transcription are based on the biological principles of the twin transcriptional loop (TTL) scenario. {\corr In this scenario, a transcribing RNAP (in pink) does not rotate, generating on each side torsional stresses in the form of torques (circular red arrows, with the sign of the supercoils indicated by $\pm$) -- note that the supercoils generated between two successive RNAPs tend to cancel each other.} 
Transcription is then {\corr usually} divided into three sub-processes: initiation, elongation and termination.  %During the elongation stage, %RNAPs translocate along the DNA and act as topological barriers. This implies that the linking numbers on each side of each RNAP are treated separately.
{\corr Elongating RNAPs are considered to act as topological barriers and} the generated torques are estimated from the corresponding supercoiling values~\cite{marko_torque_2007}. In all models, the speed of RNAPs is highest when the torques are zero and decreases as the torque values increase. Depending on the model precision,  the removal of supercoils by topoisomerases is considered either globally or locally. In the latter case, as shown in the figure, the models incorporate the distinct activities of Topo~I and DNA gyrase, which act preferentially upstream and downstream of the gene, respectively (see main text). The initiation process can be further divided into multiple stages, including the promoter binding step and the subsequent steps that lead the DNA-bound RNAP into the elongation stage. These steps encompass the formation of the open complex and the promoter escape. Elongation is typically modeled as a deterministic process, where the speed of RNAP is a function of the torque acting on it (red curve). The other stages are modeled as stochastic processes, where the corresponding rates ($\lambda_T$, $\lambda_G$, $k_b$, $k_o$, $k_e$) are often unknown and are therefore subjects of investigation (see e.g.~\cite{boulas_assessing_2023} for rates associated with topoisomerases activity). Finally, it should be noted that the question of the three-dimensional folding of DNA and its impact on the different stages of transcription is currently not considered in these models.}}
\label{fig:trans_model}
\end{figure}

%\subsection{{\corrprev Physical models of the TTL: assumptions and outcomes}}
\subsection{{\corr Physical implementation of the twin transcriptional loop (biological) model}}

%{\it Physical} models of have then been developed in the context of the {\it biological} TTL model with the aim to quantitatively rationalize transcription {\it in vivo}. 

{\corrprev The most recent {\it physical} models of the TTL thus include} the interplay between DNA, RNAPs and topoisomerases. {\corrprev In a nutshell, they consist in including altogether both stochastic and deterministic parts of the transcription process. That is, they include with different levels of precision a stochastic description of transcription initiation, a deterministic description of RNAP elongation (with the speed being a function of the torque acting on the RNAP), a deterministic description of termination and a stochastic description of the action of topoisomerases (Fig.~\ref{fig:trans_model}). They then make the assumption that any elongating RNAP behaves as a topological barrier, or that it can absorb part of the supercoiling by rotating. The motion of RNAPs is then described at } a spatial resolution of typically less than {\corrprev a few tens base pairs}. Associated torques can indeed vary dramatically as soon as the RNAP transcribes a few base pairs: for two consecutive RNAPs separated by e.g.~$100$ (500) base pairs,
%(as in RNAP convoys observed in eukaryotic cells~\cite{tantale_single-molecule_2016}),
it only requires the transcription of one (five) base pair(s) to make the supercoiling density vary by an amount of $\sim 0.01$. {\corrprev None of the models} yet include the explicit structure of DNA {\corrprev (see section~\ref{sec:simus} of the Appendix for an explanation)}. They nevertheless display a rich phenomenology that still needs to be fully understood.

More precisely, {\corrprev using these models, research groups} have endeavored to quantify the downstream accumulation of positive supercoiling and the impact of gyrase on relaxing the associated stress~\cite{sevier_mechanical_2016,ancona_transcriptional_2019,klindziuk_mechanochemical_2020}. Others have focused on the collective behavior of RNAPs~\cite{brackley_stochastic_2016,jing_how_2018,chatterjee_dna_2021,sevier_collective_2022,tripathi_dna_2022,geng_spatially_2022}. In particular, several scenarios have been proposed for the observation of non-trivial long-distance effects associated with transcription. Namely, opposite tendencies for the translocation speed of an RNAP in the presence of other RNAPs have been observed, depending on whether the promoter is active or not, with more rapid, slower respectively, translocation rates~\cite{kim_long-distance_2019}. These phenomena cannot be explained by a simple cancelation of the supercoiling between successive RNAPs {\corrprev (Fig.~\ref{fig:trans_model})}. Additional mechanisms have thus been hypothesized. These include (i) the  velocity of an RNAP that depends on the net torque that is exerted on it, i.e., the downstream torque minus the upstream torque~\cite{chatterjee_dna_2021,tripathi_dna_2022,sevier_collective_2022,geng_spatially_2022}, ii) a supercoiling stress that increases with the number of bound RNAPs~\cite{chatterjee_dna_2021}, iii) a DNA-bound {\corrprev transcription factor}, or a small DNA loop, acting as a topological barrier~\cite{chatterjee_dna_2021} and iv) a slow diffusion of the linking number~\cite{brackley_stochastic_2016,geng_spatially_2022}.

Hypothesis (i) deserves experimental testing since single-molecule experiments have thus far examined the impact of downstream and upstream torques on elongating RNAPs {\it separately}~\cite{ma_transcription_2013,ma_interplay_2014}. It also remains to be determined whether this hypothesis is consistent with an elongating RNAP's ability to act as a topological barrier. Lastly, it should be noted that quantitative modeling of transcription by separately considering downstream and upstream stalling torques is feasible~\cite{boulas_assessing_2023}.
Hypothesis (ii) echoes the observation of RNAPs that cluster when the most downstream one stalls~\cite{fujita_transcriptional_2016}, which should indeed exert a higher torsional friction. Hypothesis (iii) could be tested experimentally. Nevertheless, both experiments~\cite{leng_dividing_2011} and polymer simulations~\cite{joyeux_requirements_2020} suggest that, for DNA-bound proteins to generate a topological domain, they must embed the domain inside a loop at the very least. Finally, hypothesis (iv) was made by considering the relaxation speed of the linking number as given by the diffusion speed of plectonemes. However, both single-molecule experiments~\cite{crut_fast_2007,van_loenhout_dynamics_2012} and polymer simulations~\cite{matek_plectoneme_2015,joyeux_requirements_2020,fosado_nonequilibrium_2021,wan_two-phase_2022} have  shown that the former, which is responsible for the formation of plectonemes, is much higher than the latter. In other words, supercoiling establishment during transcription can be regarded as a quasi-static process~\cite{wan_two-phase_2022}.

%\subsection{Towards a quantitative understanding of topoisomerases}

Recently, two physical implementations of the TTL model have, for the first time, {\it separately} considered the actions of Topo~I and DNA gyrase~\cite{geng_spatially_2022,boulas_assessing_2023} (Fig.~\ref{fig:trans_model}). In particular, the model proposed in~\cite{boulas_assessing_2023} has a minimal number of parameters and, coupled with an experimental realization of the TTL model in \ecoli, has provided novel, quantitative insights into the {\corr operating mode} of topoisomerases. Specifically, it predicts that Topo~I and DNA gyrase systematically accompany gene transcription by respectively removing negative and positive turns at rates of approximately one to two (negative) supercoils per second and at least two (positive) supercoils per second. These rates are consistent with {\it in vitro} activities reported for both Topo~I~\cite{terekhova_bacterial_2012} and DNA gyrase~\cite{ashley_activities_2017}. Moreover, the model predicts that the positive {\corrprev linking numbers} introduced by Topo~I have antagonistic effects on the different stages of transcription. On the one hand, they allow the release of negative torque upstream of the RNAP so that it can properly translocate~\cite{ma_transcription_2013,ma_interplay_2014}. On the other hand, they hinder the opening of the double helix, thereby tending to repress the formation of the so-called open complex~\cite{murakami_bacterial_2003} at the initiation stage.

\subsection{{\corrprev Open problems and modeling perspectives}}

\paragraph*{\corrprev Cooperative effects between genes.}
The global nature of the conservation of the linking number (section~\ref{sec:basics}) and the quick relaxation of twist and writhe compared to the speed of supercoil generation (section~\ref{sec:1dmodel}) suggest that there is a long-range coupling of supercoiling-induced mechanical stresses that extends to topological barriers. Accordingly, changes in supercoiling around highly transcribing genes can extend up to tens of kb~\cite{visser_psoralen_2022}. Multiple experimental studies have, {\it de facto}, demonstrated that supercoiling-induced coupling affects the transcription of neighboring genes~\cite{lilley_dna_1996,opel_dna_2001}, with an impact observed at distances of several kb~\cite{hanafi_activation_2002, moulin_topological_2004}.
%Notably, this possibility was predicted by Liu and Wang in their seminal paper~\cite{liu_supercoiling_1987}.
Physical models have been developed in order to better understand these effects ~\cite{meyer_torsion-mediated_2014,yeung_biophysical_2017,johnstone_supercoiling-mediated_2022,sevier_collective_2022,geng_spatially_2022} and to understand the impact of this coupling on the organization of genomes~\cite{sobetzko_transcription-coupled_2016,geng_spatially_2022} and their possible evolution~\cite{grohens_genome-wide_2022}. So far, models have not included effects from topoisomerases, except in a very recent work~\cite{geng_spatially_2022}. Yet, the necessity to include them to understand the coupling between neighbor genes was stressed (already) 30 years ago in an analysis of the non-trivial transcriptional properties of the leucine biosynthetic operon in {\it Salmonella} Typhimurium~\cite{lilley_local_1991}. The latter has become a prototypical system of the supercoiling-based coupling of the transcription of divergent genes~\cite{lilley_dna_1996,rhee_transcriptional_1999,opel_dna_2001,hanafi_activation_2002}.

\paragraph*{\corrprev Transcriptional bursting and its time scale.} The transcription of many genes in bacteria (and eukaryotes~\cite{coulon_eukaryotic_2013}) {\corrprev has been shown to be} bursty~\cite{golding_real-time_2005}: it is governed by a non-Poissonian process of transcript production involving at least two distinct characteristic times. Namely, single-cell experiments have revealed that the dynamics of expression alternate slowly between active and inactive phases of transcription, with a characteristic time on the order of ten minutes~\cite{golding_real-time_2005,so_general_2011}. This characteristic time is much larger than those associated with the mechanisms of transcription during the active phase, whether it be the time required to transcribe the entire gene ($\sim 1$  minute) or the time between two supercoil removals by the topoisomerases (a few seconds)~\cite{boulas_assessing_2023}.  Importantly, this slow modulation of transcription depends on the activity of DNA gyrase, {\lc and the characteristic time for this modulation decreases as the concentration of DNA gyrase increases}~\cite{chong_mechanism_2014}. {\corrprev The commonly accepted rationale is the following.} RNAPs stall when the positive downstream supercoiling becomes too intense~\cite{ma_transcription_2013,ma_interplay_2014}, that is, when the supercoiling density is on the order of $+0.06$ (see section~\ref{sec:inctopo}). In the absence of DNA gyrase, transcription is therefore hindered up to the point where a DNA gyrase binds downstream and relaxes the positive supercoils. {\corrprev These observations raise important questions about the dynamics of the expression of gyrase itself. In particular, is gyrase  transcription bursty? Also, measurements in \ecoli\ have led to the conclusion that} only about 300 gyrases might be bound {\corrprev at each instant} along the genome~\cite{stracy_single-molecule_2019}, that is, one gyrase every $\sim\SI{15}{kb}$. {\corrprev While this is consistent with DNA gyrase being a limiting factor for transcription, it is not clear why the cell would actually hinder transcription elongation.}

\paragraph*{{\corr The impact of} DNA folding.}
{\corrprev So far, physical implementations of the TTL model have discarded geometrical effects associated with both the one-dimensional sequence-dependent distribution of torsional stress and the three-dimensional folding of DNA, which may impact the binding properties of RNAPs and topoisomerases.} Experimentally, the effect of local DNA folding on transcription is actually not known, except in the specific case of small DNA loops involving transcription factors~\cite{cournac_dna_2013}. Interestingly, Wang suspected that for large values of supercoiling density, folding effects would limit the accessibility of RNAP to DNA~\cite{wang_interactions_1974}. His reasoning came from the comparison of two phenomena, whose behaviors as a function of the supercoiling density were similar. Namely, on the one hand, he observed that the transcriptional activity of an RNAP, and more specifically of the core enzyme (i.e., without the ability of the RNAP to recognize specific promoters), is a non-monotonic function of supercoiling density with a maximum at values between $-0.05$ and $-0.04$. On the other hand, he observed a change in the sedimentation properties of plasmids in migration gels around $-0.035$ that he interpreted as a ``higher twisting of one double helix around the other''~\cite{wang_interactions_1974}. Years later, equilibrium studies of polymer physics models of $\SI{10}{kb}$ long supercoiled molecules confirmed this conformational effect~\cite{krajina_large-scale_2016}: when the supercoiling density decreases below $\sim -0.03$, branches become longer and tighter, which could indeed hinder accessibility to DNA. We note, here, that this structural effect could actually contribute to the systematic non-monotonic behavior of gene expression level as a function of supercoiling density observed for different promoters {\it in vitro}~\cite{pineau_what_2022}, although the ``maximal'' supercoiling values differ substantially between promoters~\cite{pineau_what_2022}. {\corrprev In all cases, models of transcription regulation involving the explicit multi-scale structuring properties of DNA remain to be developed.}
%, with a maximum value on the order of $-0.05$. Variations around this value between promoters might then come from different sentivities to supercoiling of the denaturation process involved during the formation of the open-complex at initiation~\cite{pineau_what_2022}.

%Finally, let us note that recent high-resolution (on the order of 500 bp) data  obtained in a context where transcription of a single gene is active~\cite{bignaud_transcriptional_2022} have provided crucial information on the expected properties of local DNA organization associated with transcription. Any relevant polymer physics model associated with the transcription process should thus be able to provide a rationale for the specific patterns found in these experiments, such as the presence of arrows for highly expressed genes.

\section{Supercoiling constraints and DNA replication}
\label{sec:replication}

%\subsection{Topological constrains downstream the replisome}

{\corrprev The topological problems behind and ahead of the advancing replication complex, also known as the replisome, are of a different nature. Behind, they involve the intermingling of two molecules: the replicated DNAs. Ahead, they involve a single molecule: the unreplicated DNA. Let us first recall, then, that} the DNA polymerase of mesophilic bacteria duplicates DNA at a rate of about $\SI{1000}{bp}$ per second. Composed of a large number of proteins and, hence, expected to be constrained by a high torsional friction with the surrounding biomolecules of the cytoplasm, the replisome is unlikely to rotate as quick as it introduces supercoils in DNA. Supposing no rotation at all, the replisome would thus introduce {\corrprev ahead } on the order of $100$ positive supercoils per second. Considering the presence of topological barriers located at a distance on the order of $\SI{10}{kb}$ (section~\ref{sec:transcription}), the replisome would thus make the DNA supercoiling density {\corrprev ahead} vary by an amount of $0.1$ every second  -- see below for the discussion of a rotating replisome. Since DNA replication is directly linked to the ability of bacteria to multiply, it is therefore not surprising that replisome's advancing is accompanied by a high activity of topoisomerases~\cite{khodursky_analysis_2000, mckie_dna_2021}, and more specifically {\corrprev ahead} by DNA gyrases. In this regard, high-speed single-molecule fluorescence imaging has revealed the presence in \ecoli\ of clusters containing an average of 12 gyrases (ranging from 2 to $\sim100$) and concomitant with the onset of replication~\cite{stracy_single-molecule_2019}.
%, which are used to relax the positive supercoiling generated along the yet-to-be-replicated DNA~\cite{mckie_dna_2021}.
Also, the DNA gyrase of \Bsub\ has been shown to relax up to 100 supercoils per second in single-molecule experiments~\cite{ashley_activities_2017}. In any case, the effective rate of positive supercoils removal {\it in vivo} remains unknown. We also remind that the removal of positive supercoiling by DNA gyrases is ATP-dependent with an enzymatic cycle involving the hydrolysis of two ATP molecules to remove two supercoils~\cite{wang_moving_1998}.

{\corrprev Behind the replisome}, unwinding of the two DNA strands during replication does not generate mechanical stress that would destabilize the system, as it does in transcription. The two resulting single-stranded DNA molecules are instead managed simultaneously by dedicated enzymes associated with the replication complex to build new double helices~\cite{reyes-lamothe_chromosome_2012}. However, super-structuring between replicated DNA is known to occur behind the replisome~\cite{peter_structure_1998}. To understand this phenomenon, it must be realized that although the replication complex is large, it can rotate in principle, especially because of the large torques generated {\corrprev ahead}. From a topological viewpoint, the two replicated DNA molecules extend the Watson and Crick strands of the unreplicated DNA (Fig.~\ref{fig:precat}A). The situation is thus identical to the generation of twin supercoils described in Fig.~\ref{fig:twin_scheme}, with the possibility of rotation of the unwinding machine. According to that figure, the replisome rotates in the clockwise sense, and the replicated DNA forms a right-handed superhelix (Fig.~\ref{fig:precat}A), known {\it in vivo} as precatenanes and in single-molecule experiments as braids. Importantly, precatenanes impede replicated chromosomes from diffusing away from each other. As a consequence, precatenane release is necessary for replicated chromosomes to properly segregate. Multiple lines of evidence over the last 25 years have revealed that this is primarily performed by the topoisomerase Topo~IV~\cite{zechiedrich_topoisomerase_1997,charvin_single-molecule_2003,stone_chirality_2003,wang_modulation_2008,lesterlin_sister_2012}, with additional specific contributions from Topo~III~\cite{mckie_dna_2021}.

\begin{figure}
\centering
\includegraphics[width=\linewidth]{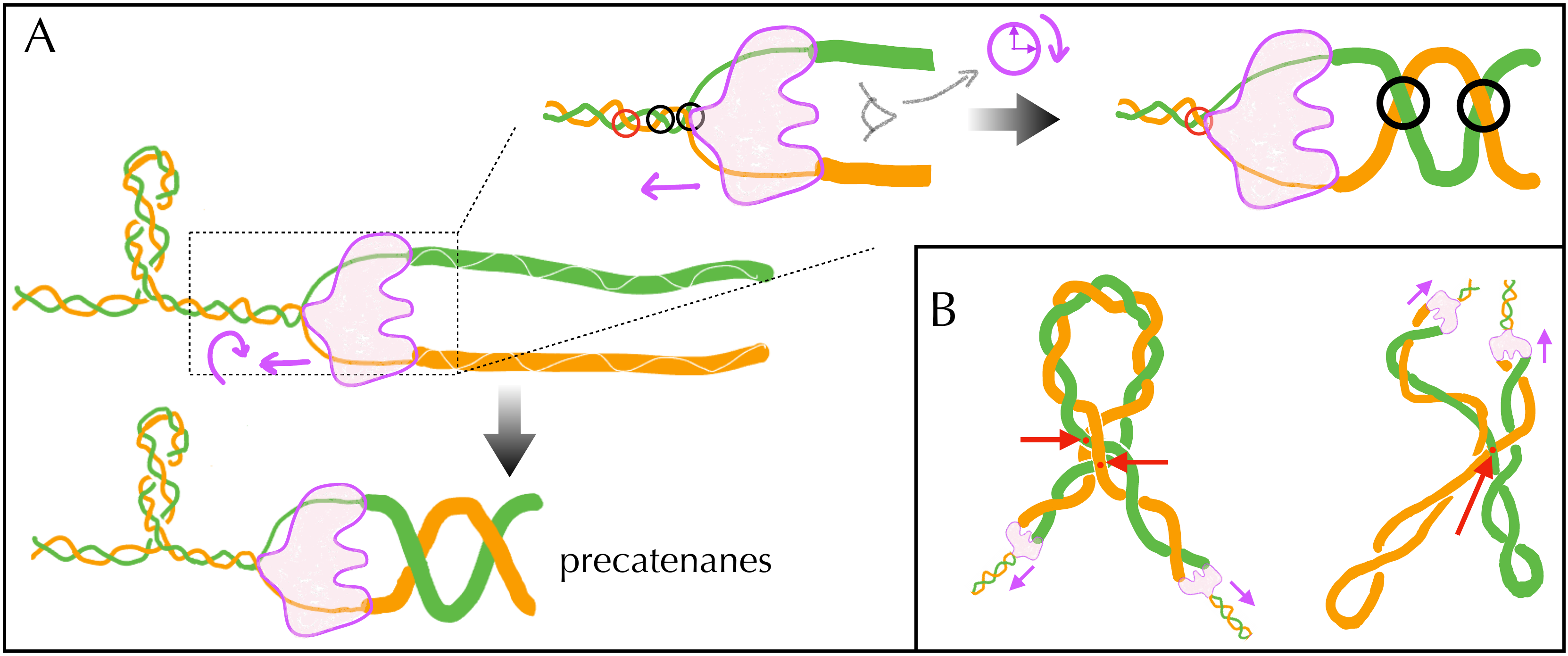}
\caption{A) Schematic representation of the formation of precatenanes during DNA replication. The replisome, indicated in light purple, moves along the unreplicated DNA double helix with the Watson and Crick strands shown in green and orange, respectively. {\corrprev Behind the replisome}, these strands give rise to two replicated molecules indicated by the thick green and orange lines, respectively. Upper panels: during the unwinding of the unreplicated DNA double helix, if the replisome rotates, it transfers the inter-strand crossings (black circles on the left) to the replicated DNA (circles on the right), which then form a superhelix (precatenane) with the same chirality. The red circle indicates a crossing that has not yet been unwound by the replisome. Lower panel: the net result of this operation is the formation of precatenanes. B) Possible conformations of precatenanes leading to left-handed crossings {\corr (red points, indicated by the arrows)}. On the left, the precatenanes buckle to form a plectonemic structure generating two left-handed crossings (adapted from~\cite{charvin_single-molecule_2003}). On the right, the intrinsic negative supercoiling of each replicated DNA leads to a left-handed crossing (adapted from~\cite{rawdon_how_2016}).}
\label{fig:precat}
\end{figure}

\subsection{{\corrprev Polymer models and the precatenane problem}}
\label{sec:precat}

%A puzzling aspect, at first sight, of the decatenation by
While precatenanes are right-handed, single-molecule experiments have shown that Topo~IV decatenates left-handed braided structures much more efficiently ~\cite{crisona_preferential_2000,charvin_single-molecule_2003,stone_chirality_2003}, raising the question of how Topo~IV would remove precatenanes {\it in vivo}. Three non-exclusive scenarios have been proposed {\corrprev in the context of a polymer physics description of precatenanes}. 
%as a consequence of the folding properties of braided molecules studied in the context of polymer physics at equilibrium (Fig.~\ref{fig:precat}B). 
First, equilibrium statistical mechanics analysis of braided molecules have shown that precatenanes, just as DNA, also buckle to form left-handed plectonemes (of precatenanes) when the density of precatenanes is sufficiently high. More precisely, defining the density of precatenanes as the ratio between the number of crossing of the two molecules, on the one hand, and the number of double helices along a single molecule, on the other hand, Marko predicted buckling to occur at a value around $0.045$~\cite{marko_supercoiled_1997}. This has been confirmed by polymer simulations of braided molecules that are stretched by pN-range forces relevant to {\it in vivo} conditions ~\cite{stone_chirality_2003,charvin_braiding_2005,forte_plectoneme_2019}. The decatenation of a right-handed precatenane could therefore occur inside a left-handed plectoneme of precatenanes (Fig.~\ref{fig:precat}B). 
%This would require in particular two enzymatic reactions instead of a single one for the left-handed precatenane, which is in accord with a reported two-fold difference in the decatenation rate between left-handed precatenanes and right-handed precatenanes in the buckling regime~\cite{charvin_single-molecule_2003}.
Second, for a number of precatenanes much below their buckling regime,
% and forces much lower than $\SI{1}{pN}$, 
%polymer simulations have shown that the angles between two braided molecules are distributed very similarly, i.e., rather broadly  with a peak close to $90^\circ$~\cite{neuman_mechanisms_2009}. In parallel, 
single-molecule experiments revealed that the chiral asymmetry in Topo~IV activity resulted from a difference in the processivity of the enzymes with respect to the chirality of the braid, with a high (low) processivity for left-handed (right-handed) precatenanes. Topo~IV could thus remove right-handed precatenanes, {\lc similar to left-handed ones, but at a slower rate}. %, but simply with a much less efficiency.
Third, polymer simulations of catenated DNA molecules at equilibrium revealed specific left-handed crossing between the two catenanes when the molecules are negatively supercoiled~\cite{rawdon_how_2016} (Fig.~\ref{fig:precat}B). Accordingly, the decatenation of sister chromatids by Topo~IV could be enhanced by negative supercoiling.

%\subsection{An inefficient process selected by Nature?}
In all cases, a puzzling question remains: why would nature select an inefficient Topo~IV decatenation activity? A common response is that Topo IV {\it should not   affect the average level of supercoiling}, with the idea that there exists some optimal value of average supercoiling~\cite{menzel_regulation_1983}. Thus, Topo IV should not intervene in the resolution of right-handed plectonemes generated upstream of RNAPs. However, this answer fails to explain why Topo~IV and DNA gyrase have overlapping activities~\cite{mckie_dna_2021,hirsch_what_2021}. Moreover, these two enzymes mainly differ at their C-terminal domain only~\cite{hirsch_what_2021}, making their inter-conversion a rather easy process from an evolutionary perspective. Instead, we surmise that the inefficiency of Topo~IV to remove right-handed plectonemes {\it allows not to interfere with the dynamics of the transcription initiation stage}. Namely, {\lc recent quantitative modeling of transcription (see section~\ref{sec:transcription} for more details) has demonstrated that transcription initiation is highly sensitive to the action of Topo~I. Topo~I indeed appears to act both as an elongation facilitator and an initiation inhibitor~\cite{boulas_assessing_2023}. The reason is that the removal of a single negative supercoil can lead to significant variations in supercoiling at the gene promoter, potentially interfering with the formation of the associated open complex. In this context, an efficient activity of Topo~IV on the right-handed plectonemes formed by negative supercoiling could significantly disrupt the delicate balance of Topo~I's activity.}

%Accordingly, the inefficiency of Topo~IV for right-handed precatenanes could thus avoid interfering with the regulation of transcription initiation, only.

\subsection{The cohesion-segregation problem: {\corrprev insights from polymer models}}

{\corrprev If not resolved, precatenanes would strongly affect the proper segregation of chromosomal loci. Actually,} replicated loci are known to remain close-by in space for at least a few minutes after the passage of the replication machinery. Considering a replication speed of $\SI{1000}{bp}$ per second, this so-called cohesion stage between chromatids {\corrprev thus concerns a post-replicative region that spread over a few hundreds kb. Details of this phenomenon} depend on multiple factors, including the type of bacteria, their growth conditions but also the timing along the cell cycle~\cite{reyes-lamothe_chromosome_2012,possoz_bacterial_2012,wang_organization_2013,kleckner_bacterial_2014,badrinarayanan_bacterial_2015}. 
%, which occurs quickly after being replicated, requires replicated chromosomes to properly drift away from each other.
%Precatenanes play a major role in this process~\cite{conin_extended_2022} as they topologically prevent replicated chromosomes to drift away from each other.
{\corrprev In all cases,} specific systems such as an increased activity of Topo~IV~\cite{el_sayyed_mapping_2016} or the action of molecular motors pulling on the replicated DNAs~\cite{bigot_ftsk_2007} are {\corrprev expected} to participate in the resolution of topological problems at the end of replication, when the density of precatenanes is a priori the highest or when only catenanes remain, i.e., when replication is finished. Nevertheless, several {\it fundamental} aspects of cohesion remain to be understood. For instance, are chromatid cohesion and precatenane formation a unique process, or can chromatids be cohesive without being topologically intermingled? Also, what are the expected respective trajectories of replicated loci once precatenanes are removed? Do they spontaneously segregate? In which directions?

In the early 2000s, the possibility of spontaneous, thermodynamically favorable segregation of intermingled sister chromatids due to the plectonemic structure of each chromatid was proposed~\cite{postow_topological_2001}. This was inspired by polymer physics modeling work showing that the probability of catenation between circular DNA and linear cyclizing DNA decreases exponentially with the supercoiling density of circular DNA~\cite{rybenkov_effect_1997} — as a consequence of a volume exclusion from the DNA compacted by the supercoiling and of the reduction of the possibilities to insert the linear DNA into the circular DNA. However, the formation of replication precatenanes is qualitatively different from this problem. A few years later, similar ideas were investigated in the context of the equilibrium statistical mechanics of catenated DNA molecules that are individually supercoiled, asking in particular the question of the amount of energy to provide to add/remove a supercoil to one chromatid {\it versus} add/remove a hypercoil from the pair of concatenated sister chromatids~\cite{martinez-robles_interplay_2009}. Two observations were discussed in particular: (i) intra-molecule negative supercoiling under the form of plectonemes make the addition of catenanes more difficult, which may hinder the production of precatenanes; (ii) segregation of the two molecules is favored by plectonemes, very likely as the result of volume-exclusion effects. Knowing that the diffusion of DNA supercoiling stresses is very fast compared to, for example, transcription rates~\cite{ivenso_simulation_2016,joyeux_requirements_2020,fosado_nonequilibrium_2021} (see section~\ref{sec:equilibrium} for details), the time scale associated with the structuring of freshly replicated DNA into plectonemes would therefore be dominated by the transcription reinitiation time (i.e., the slowest time scale). 

\subsection{Challenges ahead: out-of-equilibrium models involving long molecules}

As discussed above, the mechanisms adopted by Topo~IV and, hence, its efficiency to decatenate replicated DNA {\it in vivo} remain unknown. In particular, the spatial conformations of the precatenanes remain unknown, with at least two types of conformations that could occur (Fig.~\ref{fig:precat}B). Moreover, as far as segregation is concerned, it remains to be demonstrated that both volume-exclusion effects and entropic forces similar to those invoked to explain large-scale segregation of chromosomes ~\cite{jun_entropy-driven_2006,jun_entropy_2010} are sufficient to explain the rapid segregation of replicated chromosomes throughout the cell cycle, {\corr or whether implication of active-like segregation systems such as the ParABS system~\cite{bouet_mechanisms_2014} is required.}
%— in~\cite{jun_entropy-driven_2006}, e.g., the authors used an unrealistic outer cylinder close to the cell wall to facilitate this mechanism.

Altogether, these remarks suggest that novel theoretical studies must be performed in order to better understand the disentangling and segregation of freshly replicated chromosomes. In this regard, let us mention a minimal model that has been recently analyzed in the absence of volume exclusion effects~\cite{sevier_mechanical_2020}. It is composed of three distinct molecules (unreplicated DNA and the two copies of replicated DNA), of a converter that transforms unreplicated DNA double helices into precatenanes as well as the respective actions of DNA gyrase and Topo~IV {\corrprev ahead and behind the converter}. The objective of this work was to identify very general properties associated with the fundamental constrains on how replisomes and their associated topoisomerases process DNA. The system was analyzed in the simplifying context of a replisome that freely rotates such that the upstream and downstream torques acting on each side of it are equal. Two important results are then worth mentioning. First, in the absence of topoisomerases, it was found that the unreplicated DNA fully collapse into plectonemes before the precatenanes buckle. %, suggesting that the low precatenanes regime discussed in section~\ref{sec:precat} might be a relevant regime {\it in vivo}.
Second, to avoid this plectonemic collapse, which would trap the replisome, topoisomerases (i.e., DNA gyrase) must remove at least $\sim 1$ positive supercoil per second.

To further progress in the problem of the disentanglement and the segregation of replicated DNA molecules, it will be necessary to include the explicit structure of DNA, without which the phenomena of volume exclusion are difficult to quantify. The cost to be paid is the absence of analytical solutions and the need to resort to simulations in order to study the {\corrprev far from} equilibrium properties of the system. The numerical challenge is significant because the scales involved in the cohesion of sister chromatids (a few hundreds kilo base pairs~\cite{el_sayyed_mapping_2016}) are at least one order of magnitude greater than the typical lengths of molecules studied in Brownian dynamics (a few tens kb at most) and two orders of magnitudes greater than the lengths used in the most recent studies of precatenane-like braiding phenomena~\cite{forte_plectoneme_2019}. Methods like those used in the dynamics of rigid body~\cite{carrivain_silico_2014} thus need to be contemplated in order to improve the efficiency of the simulations.

These approaches could then give useful information in combination with data about contact frequencies between chromosomal loci, be it those allowing to differentiate sister chromatids as in the recently developed  Hi-SC2 method~\cite{espinosa_high-resolution_2020} or those resulting from standard Hi-C methods~\cite{lieberman-aiden_comprehensive_2009,le_new_2014}. Predictions should be tested in the context of topoisomerase mutants, whose effects on contact properties can be precisely quantified~\cite{conin_extended_2022}, and the activity of DNA gyrase and Topo~IV hopefully be estimated (at least for various rates of precatenane production). These approaches are also expected to provide crucial insights about how Topo~IV actually removes precatenanes {\lc {\it in vivo}} by quantifying the relative occurrence of the three mechanisms discussed in section~\ref{sec:precat} (Fig.~\ref{fig:precat}B). These models should also make it possible to validate or refute the spontaneous nature of the segregation of freshly disentangled replicated DNA.

Finally, let us mention that just as eukaryotes, bacteria contain condensins whose activity is crucial to the proper organisation and segregation of chromosomes~\cite{hirano_condensin-based_2016,gruber_SMC_2018}. Interestingly, some of the phenomena associated with the segregation of replicated chromosomes are reminiscent of the problem of the organization and segregation of mitotic chromosomes in eukaryotes~\cite{nasmyth_disseminating_2001}. Namely, Brownian dynamics simulations in the context of molecular motors extruding DNA have clarified the crucial role of condensins for chromatid segregation during prophase. The proposed mechanism relies on an effective repulsion between topologically unlinked loops~\cite{halverson_melt_2014} facilitated in this particular case by the active extrusion of intra-chromatid DNA loops by the condensins~\cite{goloborodko_compaction_2016}.
%, i.e., according to an out-of-equilibrium process acting on polymer chains.
Transposed to the problem of bacteria, these approaches offer a promising modeling framework for studying the phenomenology associated with condensins, which are known to play a fundamental role in the segregation of chromosomes~\cite{gruber_interlinked_2014,wang_smc_2014,lioy_distinct_2020} and to functionally interact with topoisomerases like Topo~IV~\cite{hayama_physical_2010,li_escherichia_2010}.

\section{Supercoiling and nucleoid formation}
\label{sec:nucleoid}

Contrary to eukaryotes, bacterial DNA is localized in a membrane-free region of the cell called the nucleoid, which was first highlighted in the 1940s -- see~\cite{robinow_bacterial_1994} for an historical review.
Recent live imaging techniques have confirmed this phenomenon, revealing more particularly the exclusion of most ribosomes from the nucleoid so that they localize at the poles of the cells (when these are cylindrical as in many bacteria, {\corrprev Fig.~\ref{fig:nucleoid}A}) -- see~\cite{chai_organization_2014} and references therein.
In \ecoli, live fluorescence imaging indicates that, independently of the time point along the cell cycle, the nucleoid occupies approximately half the main axis of the cell and the majority of the cell section, leaving only a thin layer close to the cell wall~\cite{junier_polymer_2014,wu_cell_2019}. Super-resolution techniques have reported smaller and more structured regions~\cite{spahn_super-resolution_2014}, in accord with large internal rearrangements occurring at short time scales (i.e., below 1 minute)~\cite{wu_direct_2019}. A puzzling aspect of nucleoids has concerned their specific cellular localization during the cell cycle~\cite{possoz_bacterial_2012,reyes-lamothe_chromosome_2012,wang_organization_2013,kleckner_bacterial_2014,badrinarayanan_bacterial_2015}. In \ecoli\, for instance, just after cell division the nucleoid is localized at the center of the cell. As replication proceeds, it quickly splits into two (replicated) nucleoids which localize at the quarters of the cell until cell division occurs {\corrprev (Fig.~\ref{fig:nucleoid}A)}.

The physical mechanisms responsible for nucleoid formation have fueled numerous theoretical studies (see~\cite{benza_physical_2012,joyeux_compaction_2015} for not too old reviews), with a recurring question: what is the precise role of DNA supercoiling in this matter? The latter is indeed often mentioned as contributing to DNA compaction. However, more than 30 years ago, Cozzarelli and colleagues noted that ``the extended thin form of plectonemically supercoiled DNA offers little compaction for cellular packaging, but promotes interaction between cis-acting sequence elements that may be distant in primary structure''~\cite{boles_structure_1990}. So, does supercoiling really participate in genome compaction? More specifically, is it a key factor of nucleoid formation?

\subsection{Spatial extension of a supercoiled DNA versus confinement: scaling arguments}

First and foremost, let us address the question of the spatial extension of a supercoiled circular DNA molecule under conditions of temperature and salinity equivalent to those {\it in vivo}, but without the confinement of the cell. In polymer physics, the spatial extension of a chain is quantified by its radius of gyration, i.e., the root-mean-square distance between the center of mass of the chain and each of its monomers. It is then customary to describe the large-scale behaviors of polymer chains by assessing how their radius of gyration varies with their molecular length $L$ as the latter becomes large, also known as scaling laws~\cite{de_gennes_scaling_1979}. For example, the radius of gyration of both linear and circular self-avoiding chains has been shown to scale as $L^{0.59}$~\cite{de_gennes_exponents_1972,baumgartner_statistics_1982}. Knowing that a circular chain of 30 kb has a radius of gyration on the order of $\SI{325}{nm}$ (see e.g.~\cite{walter_supercoiled_2021}), this means that a genome of $\SI{5}{Mb}$ (genomic length typical of many bacteria, including \ecoli) is predicted to have an equivalent radius of gyration of approximately $325\times(5000/30)^{0.59}\simeq\SI{6.6}{\um}$.
For comparison, an \ecoli\ cell with a length of $\SI{2}{\um}$ and a radius of $\SI{0.5}{\um}$ has a much smaller equivalent gyration radius of $\simeq \SI{0.67}{\um}$.
In particular, the volume of the bacterium is $(6.6/0.67)^3 \approx1000$ times smaller than the typical volume spanned by its thermally fluctuating, unconstrained circular DNA.
% which is much larger than the radius of bacterial cells ($\simeq \SI{0.5}{\um}$ in \ecoli).

As discussed in section~\ref{sec:equilibrium}, a supercoiled circular DNA molecule adopts tree-like conformations, which is expected to strongly affect these results. Interestingly, by neglecting the details of this tree, such as the distribution of branch sizes, one can estimate the corresponding scaling law. Namely, scaling arguments~\cite{daoud_conformation_1981,khokhlov_array_of_obstacles_1985,gutin_annealed_rings_1993,everaers_flory_review_2017}, analytical approaches~\cite{parisi_sourlas_1981} and numerical simulations~\cite{rensburg_nonlocal_1992,cui_chen_tree_MC_1996,rosa_tree_simulation_2016,rosa_beyond_2017} have shown that the radius of gyration of self-avoiding trees scales as $L^{0.5}$. Knowing that a circular chain of $\SI{30}{kb}$ has a radius of gyration on the order of $\SI{200}{nm}$ in the plectonemic phase~\cite{walter_supercoiled_2021}, the corresponding extension for our $\SI{5}{Mb}$ long bacterial genome is equal to $200\times(5000/30)^{0.5}\simeq\SI{2.6}{\um}$, in accord with more precise calculation~\cite{cunha_polymer-mediated_2001}. While this is a significant reduction compared to topologically unconstrained circular DNA, the corresponding volume is still 60 times larger than the volume of the cells. 

\subsection{Adding (large) molecular crowders: segregative phase separation}

The scaling arguments outlined above suggest that supercoiling alone cannot account for the formation of the nucleoid, as the unconfined resulting tree would occupy a much larger volume than the bacterial cell itself. Furthermore, these arguments do not address the specific issue of the nucleoid's location within the cell, whether it is located at the center or the quarters of the cell. One significant factor missing from these arguments is the physical nature of the cytoplasm and the potential for microcompartmentalization caused by liquid-liquid phase separation~\cite{walter_brooks_phase_separation_from_crowding_1995,Hyman_LLPS_2014}.
In particular, the nucleoid might form due to depletion interactions~\cite{Asakura_Oosawa_1954,Asakura_Oosawa_1958,Lekkerkerker2011} between the bacterial DNA and ``crowders'' contained in the cellular solvent in which it is immersed~\cite{odijk_osmotic_1998,mondal_entropy-based_2011,joyeux_bacterial_2020} {\corrprev (Fig.~\ref{fig:nucleoid}A)}. 

Molecular crowders are typically identified with small, $\sim\SI{5}{nm}$ sized proteins, which are present in the cytoplasm in large concentrations.
Their presence affects the mobility of biomolecules, protein folding and stability, and the association of macromolecules with each other~\cite{van_den_Berg_crowding_homeostasis_2017} as well as the structure and stability of DNA~\cite{Miyoshi_crowding_DNA_2008}. However, the formation of the nucleoid might owe more to the presence of larger crowders like ribosomes or  polysomes (small polymers of ribosomes connected by the messenger RNA they are sitting on)~\cite{mondal_entropy-based_2011,joyeux_bacterial_2020} .

Quite generally,  large structures (like spheres, plates or rods) can be pushed together by smaller molecules, as this reduces the total volume inaccessible to the crowders and hence maximizes their translational entropy and the total disorder in the system.  In a nutshell, the compressing forces arise because the osmotic pressure of crowders in open spaces cannot be balanced due to their absence from inaccessible spaces.
Depletion interactions are particularly effective for rod-like particles, where nematic ordering can arise for similar reasons~\cite{frenkel_onsager_revisited_1987}
and mixtures of spheres and rod display a rich phase diagram as a function of their relative size and concentration (Fig.~\ref{fig:nucleoid}B)~\cite{adams_separation_of_rods_and_spheres_1998,dogic_layering_for_spheres_and_rods_2000,urakami_spheres_and_rods_2003}. Of special interest for this review are the implications of DNA supercoiling and, in particular, the importance of the length and the stiffness of the rod-like plectonemic regions in between branch points.

\begin{figure}[t]
\floatbox[{\capbeside\thisfloatsetup{capbesideposition={right,top},capbesidewidth=0.5\linewidth}}]{figure}[\FBwidth]
{\caption{{\corrprev Potential mechanisms involved in the formation of the bacterial nucleoid. A) Left: {\lc In rod-shaped bacteria such as \ecoli}, most DNA is localized in the nucleoid (grey area) at the center of the cell, while ribosomes tend to concentrate at the poles (red area). In slowly growing \ecoli\ cells, following cell division the origin of replication (in green) is positioned at the center of the cell. During the cell cycle, the replicated origins rapidly segregate towards the quarters of the cell and remain there until cell division occurs. Right: Zooming in on the periphery of the nucleoid, the DNA (plectonemic structures in black) exhibits a tendency to separate from ribosomes (in red). Additionally, polysomes may form when multiple ribosomes simultaneously translate the same messenger RNA (red dashed ellipse). B) In this context, the nucleoid has been proposed to arise from a phase separation process between spherical (ribosomes) and rod-like (plectonemes) structures. Depending on the relative sizes and concentrations of molecular species, at least four scenarios can arise: a well-mixed solution (left) and three distinct phases where rods and spheres undergo demixing (adapted from~\cite{urakami_spheres_and_rods_2003}). The problem for which the rods would be disposed along a polymer chain similar to that expected for the large scale internal structuring of chromosomes (Fig.~\ref{fig:scaleup}) remains open.}}\label{fig:nucleoid}}
{\includegraphics[width=\linewidth]{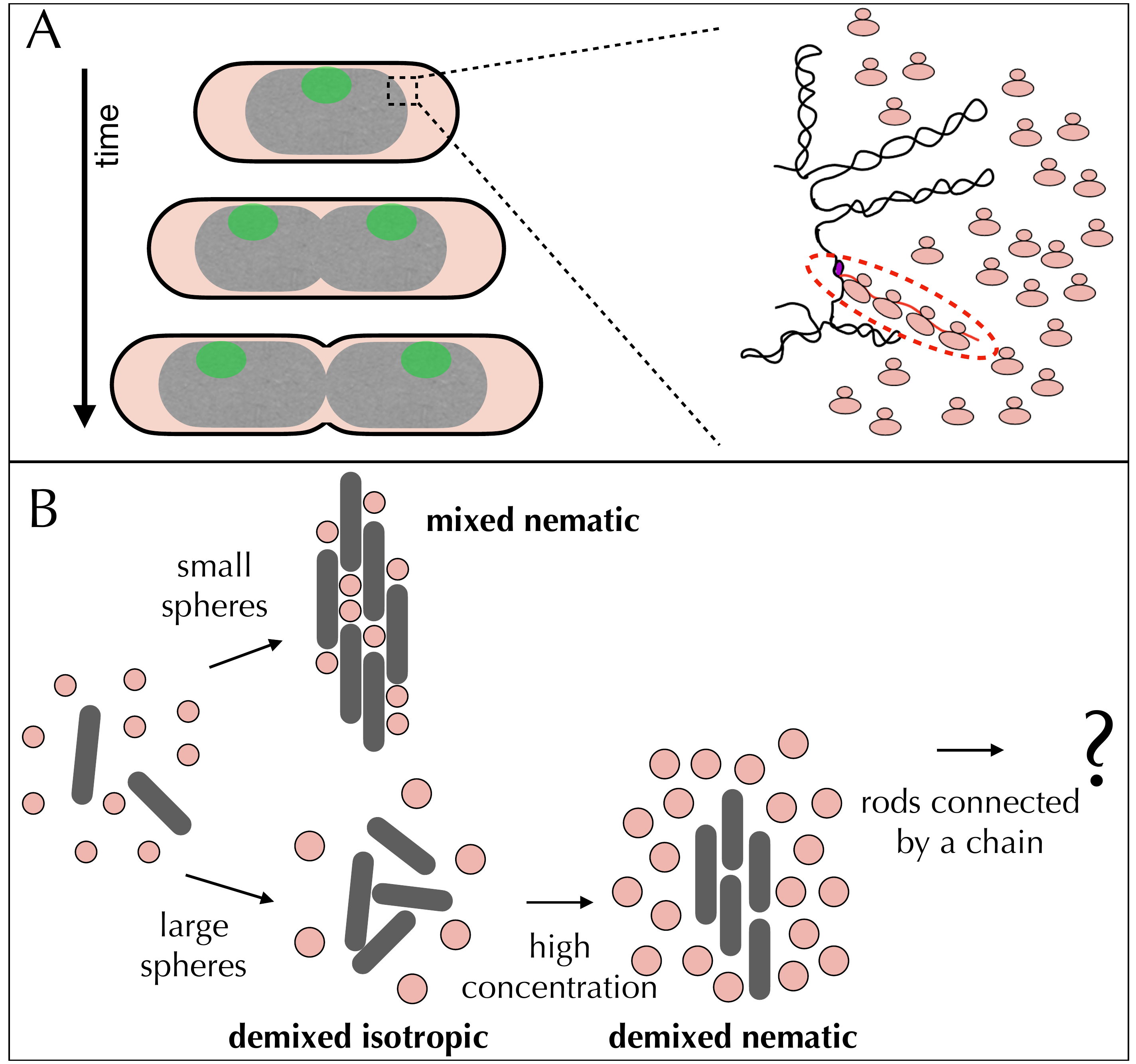}}
\end{figure}

Crowding-induced segregation of plectonemic DNA into a nucleoid was first invoked in 1998 for physiological concentrations of small proteins~\cite{odijk_osmotic_1998}. However, the equally predicted nematic ordering of the supercoiled DNA has never been observed. Instead, supercoiled DNA appears to mix with small crowders in {\it in vitro} experiments~\cite{gupta_compaction_2017} and even with $\SI{15}{nm}$ crowders in Brownian dynamics simulations~\cite{joyeux_bacterial_2020}. In 2011, a cell-scale model suggested that plectonemic DNA and polysomes undergo segregative phase separation, resulting in a similar phenomenon to that of the nucleoid in \ecoli: the plectonemes do not exhibit nematic ordering, and the chromosome tends to be localized in the center of the cell, with the polysomes congregating at the poles and in a thin layer between the chromosome and the cell walls~\cite{mondal_entropy-based_2011}. The simulation assumed that the crowders had a diameter of $\SI{20}{nm}$, slightly larger than in~\cite{joyeux_bacterial_2020}, and modeled the bacterial chromosome as a self-avoiding random tree with braided supercoiled DNA branches, approximately $\SI{1}{kb}$ ($\SI{200}{nm}$) in size. Notably, the branches were assumed to be {\it straight, i.e., very stiff}. In this context, the absence of nematic ordering is consistent with previous findings~\cite{urakami_spheres_and_rods_2003} where mixtures of rods and spheres with similar diameters exhibited such a phenomenology for a certain concentration of the spheres {\corrprev (Fig.~\ref{fig:nucleoid}B)}. Interestingly, in a mixture of rods of spheres of different sizes, there also exists a regime where the smallest spheres freely mixed with the rods, while the largest spheres may induce the nematic ordering anticipated in~\cite{odijk_osmotic_1998}.

Interestingly, in the model of~\cite{mondal_entropy-based_2011}, the chromosome avoids the cell wall to preserve the orientational entropy of the stiff plectonemes. Even more remarkably, {\corr the model} predicted that once activated, through the physical coupling of transcription and translation (section~\ref{sec:transcription}), transcribed genes should migrate to the surface of the nucleoid. This was experimentally demonstrated a few years later using live cell super-resolution imaging~\cite{stracy_live-cell_2015}. An important question nevertheless remains: are straight plectonemes of $\SI{200}{nm}$ in size (as used in the model) biologically relevant, knowing that their persistence length is on the order of $\SI{100}{nm}$, i.e., that they can actually bend rather easily below $\SI{200}{nm}$? Should one interpret the good agreement between modeling and experimental observations as indirect evidence for the association of plectonemes into stiffer bundles? If not, how would this affect the observed nucleoid phenomenology? Which additional ingredient would be necessary to add in this case? A physical coupling between part of the chromosome and the polysomes to include active genes?

Finally, recent visualization of the nucleoid in single non-dividing cells with a growing membrane have shown that a single nucleoid diffuses slowly compared to its internal dynamics, regardless of the cell length. Additionally, it diffuses slowly enough compared to the rate of cell division that it remains at the center of the cell, even when the cell becomes artificially very long~\cite{wu_cell_2019}. To understand this effect, let us first mention that experiments of \ecoli\ chromosome micromanipulation have shown that it behaves {\it in vivo} like a highly compressed spring, meaning that the pressure exerted by the cytoplasm is much greater than that required to fit the chromosome inside the cell~\cite{pelletier_physical_2012}. Thus, in a first approximation, the chromosome can be seen as a double-piston for which the cytoplasm exerts strong pressure on each side~\cite{wu_cell_2019}. As the volumes on each side of this piston contain on average equal amounts of proteins, they exert comparable pressure. Nevertheless, the proteins can pass from one side to the other through e.g.~the thin layer between the chromosome and the cell wall. The question then is to know the time scale associated with these fluctuations. An interesting insight comes from the modeling work accompanying the experiments of~\cite{wu_cell_2019}. Namely, the authors implemented molecular dynamics simulations of a brushed polymer, i.e., of a polymer composed of a (rather stiff) ring to which loops, which could be plectonemes, are attached (Fig.~\ref{fig:scaleup}). This brushed polymer was then immersed in a medium mimicking a cytoplasm crowded by ribosomes. Their results then support the idea that under these conditions, the chromosome diffuses slowly~\cite{wu_cell_2019}, very likely because of rare exchanges of ribosomes between the two sides of the polymer. Accordingly, in the presence of two nucleoids, they showed that the continuous addition of ribosomes distributed equally on either side of the corresponding polymers led to a cellular arrangement with two nucleoids located at the quarters of cells.

{\corr In all cases, the same models explaining the formation of the nucleoid as a result of crowding-induced segregation between DNA and ribosomes/polysomes should be able to account for the absence of segregation observed in a few bacteria~\cite{choi_relation_1996}. This absence of segregation could correspond to a mixing phase within the space of relevant parameters, such as ribosome density, size, and rigidity of plectonemes. Additionally, it is important to consider that other mechanisms might also play significant roles in the formation of the nucleoid~\cite{joyeux_compaction_2015}, including the bridging effect of certain nucleoid-associated proteins~\cite{dillon_bacterial_2010} or the dynamic formation of loops by bacterial condensins~\cite{hirano_condensin-based_2016,gruber_SMC_2018}.}

\section{Scaling up models of supercoiled DNA}
\label{sec:scaleup}

\subsection{On trees and bottle brushes}

One way to scale up polymer models of supercoiled DNA is to consider that braided structures such as plectonemes behave like self-avoiding linear polymers with, for example, an equivalent diameter of the order of {\corrprev $\sim \SI{10}{nm}$} for $\sigma=-0.05$~\cite{boles_structure_1990}. It then becomes possible to use a classical linear chain modeling without topological constraints (such as a wormlike chain) to address the problem of large-scale polymer folding. In this case, one must nevertheless ask how these pieces of linear chain are connected together. Two possibilities have particularly caught the attention of researchers: tree structures and bottle brush organizations (Fig.~\ref{fig:scaleup}).

The tree-like structures are observed in vitro without the action of enzymes and proteins acting on DNA~\cite{boles_structure_1990} as well as in polymer simulations (see e.g.~\cite{krajina_large-scale_2016,walter_supercoiled_2021} for molecules above $\SI{30}{kb}$ in length). Tree-like models are therefore good candidates to predict behaviors at large scales, i.e.~when the details of the trees, such as the length of their branches, do not have an impact on the studied properties -- see~\cite{daoud_conformation_1981,everaers_flory_review_2017} and references therein for the physics of trees. A characteristic example is the behavior of the average contact frequency between loci as a function of genomic distance ($s$), generally called the ``contact law'' and denoted by $P(s)$~\cite{mirny_fractal_2011}. Specifically, in situations of {\it high polymer concentrations}, simulations of trees lead to contact laws of the form $P(s)\sim s^{-1.1}$~\cite{rosa_conformational_2019}. Interestingly, this law seems to be compatible with observations in very different bacteria, namely \Ccres~\cite{le_high-resolution_2013}, \ecoli~\cite{lioy_multiscale_2018}, \Pseudo~\cite{varoquaux_computational_2022}, or {\it Streptomyces}~\cite{lioy_dynamics_2021}.
% A thorough study of the relevance, or not, of tree-like models for contact properties at various scales is ongoing~[Ghobadpour et al., in prep]. 

Remarkably, $P(s)\sim s^{-1.1}$ is actually also compatible with large-scale contact properties of chromosomal loci in  several eukaryotes such as Human~\cite{lieberman-aiden_comprehensive_2009}. While it is tempting to ascribe this here as well to DNA supercoiling known to occur in eukaryotes~\cite{Giaever_supercoiling_in_eukaryotes_1988,Corless_effects_2016,Corless_investigating_2017},
the commonly invoked explanations of crumpling~\cite{cremer_territories_2001,grosberg_crumpled_DNA_1993,rosa_interphase_chromosomes_2008,lieberman-aiden_comprehensive_2009,mirny_fractal_2011,halverson_melt_2014}
and active loop extrusion~\cite{goloborodko_compaction_2016} also lead to double-folded branching structures~\cite{khokhlov_array_of_obstacles_1985,grosberg_ring_melt_2014,rosa_crumpling_2014,rosa_tree_melt_2016,everaers_flory_review_2017,rosa_conformational_2019}.
Note, also, that the high concentration nature of the polymers for bacteria is a consequence of an {\it in vivo} concentration to be considered that is not that of DNA, which is a few percent, but that of plectonemes bound by multiple proteins. Namely, a rough calculation assuming beads of diameter $\SI{30}{nm}$ (consisting of $\sim 10$ nm in diameter and $\sim 20$ nm of protein complexes) with $\SI{200}{bp}$ per bead results in a volumetric fraction of beads of approximately $0.25$ for a $\SI{5}{Mb}$ genome folded within a nucleoid with a cross-section of $\SI{800}{nm}$ and a length of $\SI{1}{\um}$.

Regarding the organization in bottle brush, it should be mentioned firstly that based on biochemical and biophysical analyses of nucleoids extracted from cells, a rosette structure was predicted 50 years ago for the \ecoli\ chromosome~\cite{worcel_structure_1972}. In this structure, long plectonemes (of approximately $\SI{100}{kb}$) emanate from a central core made of proteins and RNA. This structure was later confirmed by electron microscopy observations of nucleoids extracted from cells~\cite{kavenoff_electron_1976}. However, {\it in vivo} evidence for such a rosette structure has remained elusive so far. Interestingly, recent experiments in which DNA replication and cell growth were decoupled led to widened cell geometries inside which a toroidal geometry of the circular chromosome of \ecoli\ could be clearly identified~\cite{wu_direct_2019}. This structure is compatible with a circular bottle brush polymer model, which is a polymer model made of a circular backbone along which plectonemes are attached (Fig.~\ref{fig:scaleup}B).

Interestingly, the chromosome of \ccres\ has been modeled using such a bottle brush polymer model in order to provide a rationale for the patterns observed in the first bacterial Hi-C data produced 10 years ago~\cite{le_high-resolution_2013}. In this model, the plectonemes were stochastic structures whose length was adjusted along with $5$ other parameters (such as the stiffness of the plectonemes or their distance along the backbone) to reproduce the Hi-C data. Interestingly, the plectonemes in the obtained model had an average length of 15 kb, which is compatible with the length of topologically independent domains predicted to partition bacterial genomes (see section~\ref{sec:3dmodel}). Furthermore, the introduction of plectoneme-free zones blocking the diffusion of plectonemes allowed for the reproduction of the phenomenology of so-called chromosome interaction domains, or CIDs, inside which interaction between any pair of loci is enhanced compared with external loci located at a similar genomic distance~\cite{le_high-resolution_2013}. Finally, some of the large-scale conformations of this model adopted a loose helix conformation, a property that has been reported for the \ecoli\ chromosome~\cite{hadizadeh_yazdi_variation_2012,fisher_four-dimensional_2013}. This is in contrast to the early data-driven ``models'' of \ccres\ chromosomes presenting a marked helix~\cite{umbarger_three-dimensional_2011} but whose origin was not physical, as demonstrated in a more physical version of these models by including the fundamental concept of entropy~\cite{messelink_learning_2021}. Note also that several theoretical studies have been carried out on these bottle brushes, highlighting helical structures in a regime where the backbone persistence length is at least of the order of the cell diameter~\cite{chaudhuri_spontaneous_2012,jung_confinement_2019}. The relevance of this hypothesis for {\it in vivo} situations remains to be demonstrated. Finally, it is noteworthy that the bottle brush polymer model, which was developed for \ccres~\cite{le_high-resolution_2013}, has recently inspired a data-driven approach aimed at creating a three-dimensional representation of the current knowledge on the structuring of bacterial chromosomes~\cite{hacker_features_2017}.

\begin{figure}[t]
\floatbox[{\capbeside\thisfloatsetup{capbesideposition={right,top},capbesidewidth=0.35\linewidth}}]{figure}[\FBwidth]
{\caption{Two types of polymer models including the effects of supercoiling can be contemplated to study the large-scale structure of bacterial chromosomes: A) tree-like models where plectonemes are abstracted by simple linear branches (right panel). B) bottle brush models where plectonemes are attached along a ring or backbone, indicated in blue. This model is therefore composed of two {\it a priori} independent entities and, hence, is more complex than the tree-like model. At large scales, the details of these entities can nevertheless be discarded (right panel). It should be noted that if the bottle brush structure is relevant {\it in vivo}, as suggested by chromosome visualization data in \ecoli~\cite{wu_direct_2019}, the mechanisms of its formation remain an open question.}\label{fig:scaleup}}
{\includegraphics[width=\linewidth]{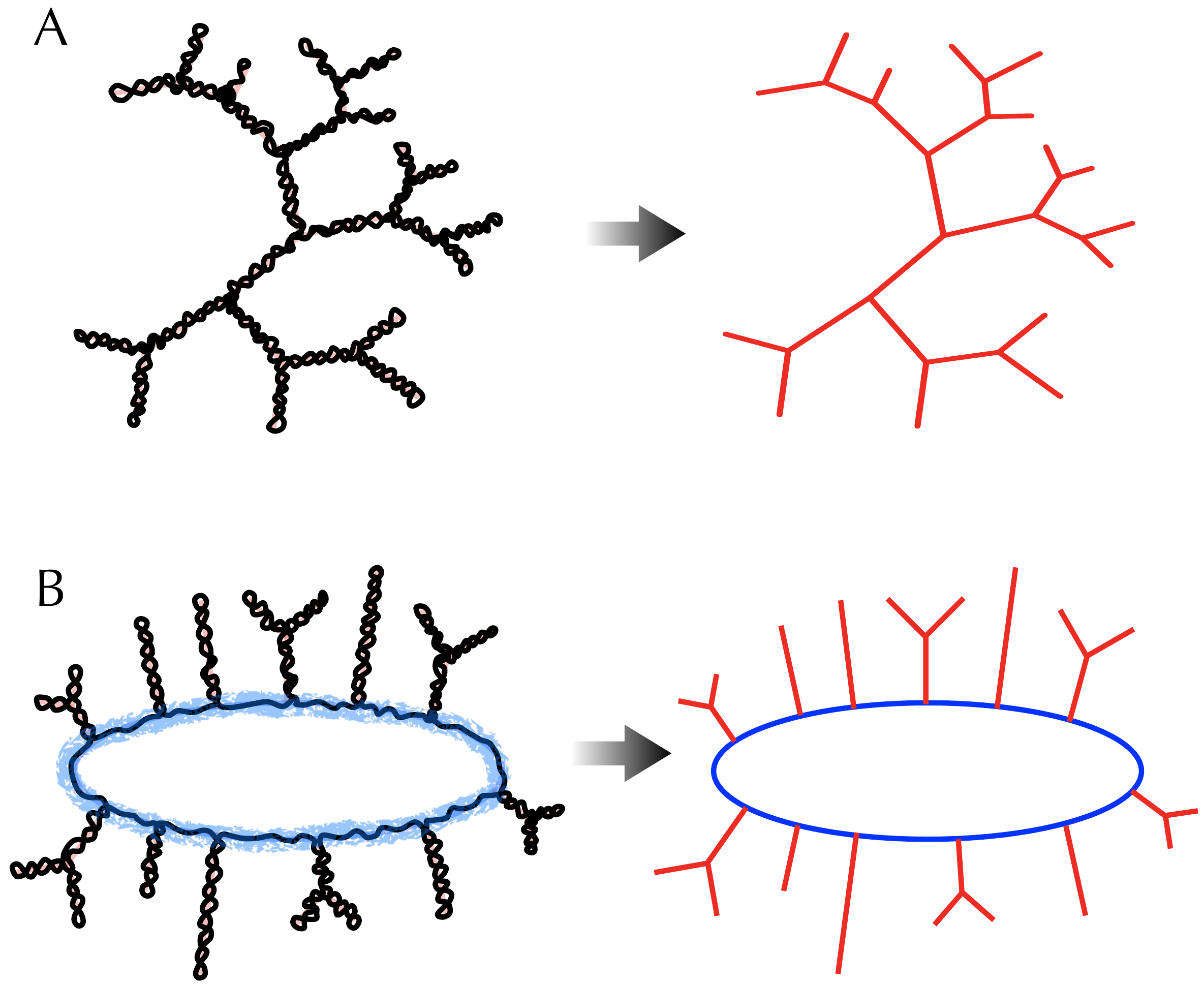}}
\end{figure}

\subsection{On-lattice models}

The simulation of tree-like models is commonly performed on a lattice. Lattice simulations are preferred due to the ease of managing discrete elementary movements as compared to continuous movements involved in off-lattice approaches. This leads to higher efficiency of lattice simulations. In fact, lattice simulations are particularly suitable when the properties under study occur on a much larger scale than the lattice mesh, i.e., when the properties studied do not depend on the geometric parameters of the lattice. The possibility of performing non-local movements, such as cutting a branch at one point and randomly reintroducing it at another point~\cite{SeitzKlein1981,rensburg_nonlocal_1992,rosa_tree_simulation_2016,rosa_tree_melt_2016}, or the exchange sections of overlapping chains~\cite{Theodorou2002PhysRevL} or trees are particularly effective on a lattice~\cite{svaneborg2016multiscale,svaneborg2023multiscale} and allow to reach thermodynamic equilibrium very efficiently. Elastic chain methods on the lattice are also very effective for exploring polymer dynamics in situations where the polymer concentration is very high~\cite{EvansEdwards1981-Part1,Barkema.Book1999,Barkema2003The.JofCh.Phy,Kolb2009Macromolecules,Schram2013TheJ.ofCh.Ph}. In a nutshell, the principle is based on the ability to redistribute monomers along a given spatial conformation, with several consecutive monomers being able to overlap. This then allows for the exploration of new conformations that would be inaccessible without this prior redistribution. Finally, simulation techniques can be adopted to reproduce realistic dynamic properties of polymers~\cite{ghobadpour_monte_2021}, as well as simulate active processes such as the action of condensins~\cite{miermans_lattice_2020}.

To our knowledge, no work has reported on the properties of a lattice-based physical model that covers the multiple scales of the bacterial chromosome. However, it is worth mentioning that a computational method has been developed to efficiently construct lattice-based conformations of a bottle brush polymer with a backbone to which plectonemes are attached~\cite{goodsell_lattice_2018}. The plectoneme modeling used in this study bears resemblance to the double-folding lattice polymer models, where linear chains fold back on themselves to form overlapping double-chain structures. In this regard, we believe that the range of methods developed in this specific area of polymer physics  should allow for a precise and quantitative analysis of the physical nature of bacterial chromosomes.
{\corr Specifically, models should be capable of explaining both contact properties and the spatial positioning of loci identified through fluorescence visualization~\cite{espeli_dna_2008,hong_caulobacter_2013}. Interestingly, it seems that within models, the latter naturally arises from the former when forcing the localization of only a few specific loci, such as those associated with the origin and terminus of replication~\cite{messelink_learning_2021}. In this context,} an important open question to us is the following: is it possible to distinguish between tree-like and bottle brush-type phenomenologies based solely on contact properties between loci as provided by Hi-C data, knowing that the latter can be generated in principle for any type of bacteria cultivable in the laboratory~\cite{marbouty_generation_2017}?

\section{Concluding remarks}

%\paragraph*{Caveats.} The situation just described also 

{\corrprev In this review, we have discussed various models of bacterial supercoiled DNA, which differ in the scales they describe and the types of processes involved. We have specifically distinguished between structuring phenomena that can be described using thermodynamic equilibrium approaches and phenomena that operate far from equilibrium, such as gene transcription or DNA replication.

One fundamental question, which is expected to gain increasing importance, especially within the field of systems biology, is whether it is possible to develop a physically-grounded unified framework that integrates these different modeling perspectives. The challenge lies in developing a multi-scale model of biophysical phenomena, wherein identifying a hierarchy of mechanisms, if it indeed exists, can be extremely difficult. This problem of developing a hierarchy of descriptions is already a challenge in the study of physical matter, particularly in the context of numerical simulation~\cite{steinhauser2017computational}. In the case of biological matter, and more specifically in the field of Chromosome Biology, this problem is even more central. A comprehensive understanding of phenomena indeed requires, in principle, considering scales ranging from the base pair level to the cellular organization of chromosomes.

Next, the discussed} models often neglects the interactions of DNA with proteins and molecular machines, as well as with all the small molecules and ions that make up the cytoplasm. Although coarse-grained models of (supercoiled) DNA have proven successful in single-molecule experiments, it is therefore reasonable to question how well these models capture the behavior of DNA in a living cell. Even more worrying for a rational approach to the phenomena at play, proteins and molecular machines often have their own specificity that arises from the molecular tinkering induced by natural selection~\cite{jacob_evolution_1977}. Many of their properties therefore escape the universality feature of physical phenomena.

%\paragraph*{Statistical universality and ubiquitous phenomena.}
The relevance of coarse-grained models {\corrprev nevertheless} arises from two realities. First, in many situations, the conditions are equivalent to those of a system with a large number of particles or in the limit of a very large size of the entities involved. In this case, statistical physics approaches become relevant. For example, while the plectonemic structure of supercoiled DNA may be a simplifying average view of the dynamics of DNA interacting with many proteins and molecular machines, this average behavior becomes probably relevant at much larger scales, such as the chromosome, and a tree-like description of the problem should capture a good part of the associated phenomena. Second, evolutionary conserved phenomena are often associated with generic physical properties~\cite{junier_conserved_2014}. For instance, the double helix nature of DNA necessarily creates topological problems that require dedicated enzymes to resolve. This has two consequences: first, topoisomerases are ubiquitous in living organisms; and second, generic physical models for handling topological constraints can be considered, regardless of the mechanisms involved. Variations in behavior between bacteria should then reflect the possible range of physiologically relevant parameters. In all cases, proposed physical models should be evaluated not only for their descriptive (i.e., postdictive) capacity but also, and perhaps most importantly, for their predictive power.

\appendix
\section{A crash introduction to physical modeling of supercoiling-related phenomena}
\label{sec:app_models}

{\corrprev In this appendix, we aim to provide a concise and accessible introduction to biophysical models of supercoiled DNA. To accomplish this, we first introduce the concept of coarse-graining, which determines the level of detail captured by models. Next, we delve into the principles of Statistical Mechanics, a framework widely used by biophysicists. As an example, we discuss the simplest model for studying the folding properties of supercoiled DNA, known as the rod-like chain model. We then distinguish three main situations, depending whether models are studied or defined i) at thermodynamics equilibrium, ii) out of equilibrium or iii) far from equilibrium. Additionally, we discuss the power of phenomenological approaches, which allow to capture system properties in an approximative, yet often quantitative way. Finally, we discuss the Monte Carlo and Brownian Dynamics methods that are commonly used to simulate the folding of DNA in the context of these models.

\subsection{Coarse-graining level: deciding which details to drop}
\label{sec:coarse}

The first step in constructing a physical model of DNA involves determining the level of approximation, known as the coarse-graining level, which defines the spatial and temporal scales below which structural and mechanistic details are discarded. For instance, studying the effects of DNA supercoiling does not require explicit consideration of the quantum physics of atoms and chemical bonds. The most detailed models actually operate at the resolution of individual nucleotides~\cite{harris_modelling_2006,ouldridge_structural_2011,manghi_physics_2016}. Their applicability is nevertheless limited to relatively small molecules due to the time-consuming nature of simulations involved. In this review, we discuss coarse-graining approaches above the double helix, typically spanning tens of base pairs or more. These models allow investigation of properties at scales ranging from kilobase pairs to megabase pairs. Importantly, models with resolution above the double helix neglect the specific structure of the double helix itself. As a consequence, they necessitate the inclusion of an effective treatment for conserving the linking number (see below).

%It should be also noted that in specific cases, such as predicting the distribution of torsional stresses along genomes (section~\ref{sec:1dmodel}) or studying the interplay between transcription and DNA supercoiling (section~\ref{sec:transcription}), models with resolution at the single base pair level are required. However, compared to the detailed structural models discussed in~\cite{harris_modelling_2006,ouldridge_structural_2011,manghi_physics_2016}, these models incorporate additional simplifications, such as neglecting the writhe or assuming instantaneous equilibration of DNA properties.

\subsection{Statistical mechanics: the example of the rod-like chain model}
\label{sec:statmec}

Once the coarse-graining level is chosen, a model of DNA can be constructed using the principles of Statistical Mechanics. This branch of physics focuses on predicting the macroscopic properties of systems comprised of microscopic entities. The specific internal and interaction properties of these entities determine the parameters of the models. A classic example related to DNA supercoiling phenomena is the rod-like chain model~\cite{vologodskii_conformational_1992,bouchiat_elastic_2000}. In this model (Fig.~\ref{fig:models}A), DNA is represented as a series of articulated rigid segments, where the relative orientation of each segment is constrained by two parameters: the bending and torsional moduli. These parameters quantify the resistance of DNA to bending and torsion, respectively. Typically, the associated energy costs are expressed as quadratic functions of the differences in tangent and, respectively, angular orientations between two adjacent segments (Fig.~\ref{fig:models}A), with a proportionality constant specified by the moduli. Next, the conservation of the linking number can be implemented ``locally'' using the ``parallel transport'' approach~\cite{bergou_discrete_2008}, which imposes a specific analytical form for the relationship between twist and the relative orientation of contiguous segments -- the relative orientation of the Euler frames associated with each segment, to be more precise~\cite{carrivain_silico_2014,lepage_polymer_2019} (Fig.~\ref{fig:models}A). It is worth noting that the twist has actually often been defined using the very angles characterizing the Euler frame, which do not satisfy the condition of ``parallel transport''. However, in this case, the linking number only exhibits slight fluctuations around its expected value, making this Euler-based definition a valid practical approximation~\cite{vologodskii_conformational_1994}.

A common macroscopic property explored in the rod-like model is the spatial extension of the molecule and how it relates to the supercoiling density. In this regard, it is important to note that this model allows segments to overlap in space. More realistic models can be constructed by incorporating electrostatic repulsions between segments, leading to the self-avoiding rod-like chain model~\cite{vologodskii_conformational_1992}. Additionally, alternative forms of DNA can be considered~\cite{manghi_coupling_2009,efremov_transfer-matrix_2016,lepage_modeling_2017,lepage_polymer_2019}. By utilizing these refined models, macroscopic properties like the fraction of super-structuring or the fraction of denatured DNA monomers can be examined. Finally, if DNA/RNA polymerases and topoisomerases are incorporated into the model, the system becomes more complex, requiring the consideration of additional parameters to fully describe it. These include, for example, the rates at which topoisomerases remove supercoils or the speed at which RNA/DNA polymerases translocate along DNA. Additional macroscopic properties relevant to the functioning of bacteria can then be examined such as the production rate of RNA transcripts.

\subsection{Equilibrium statistical mechanics and phenomenological approaches}
\label{sec:eq_phen}

The simplest scenario for investigating properties in models like the rod-like chain model is when all system changes can be attributed solely to thermal energy exchanges with the solvent (cytoplasm).%, excluding ATP-consuming transitions.
This assumption establishes the framework of ``equilibrium statistical mechanics'', which states that the probability of any molecular configuration of the chain is proportional, at long times, to the Boltzmann weights: $\exp[-E/k_BT]$. $E$ represents the energy of the configuration, reflecting bending and torsional costs, $k_B$ is the Boltzmann constant, and $T$ is the cytoplasm's temperature. In certain cases, such as in the regime of low supercoiling where plectonemes of the actual DNA molecule are not distinguishable, it becomes possible to precisely calculate the average and variance of properties like the spatial distance of the DNA chain~\cite{bouchiat_elasticity_1998,bouchiat_elastic_2000}. For example, in 1998, Bouchiat and Mezard presented a semi-analytical solution (involving the numerical solution of a system of two equations) for the average spatial extension of the continuous version of the rod-like chain model, as a function of supercoiling density and the stretching force acting on it~\cite{bouchiat_elasticity_1998}. Their results demonstrated excellent agreement with those obtained from single molecule experiments~\cite{bouchiat_elasticity_1998,bouchiat_elastic_2000,strick_stretching_2003}.

In the most general case, deriving exact or nearly exact solutions for equilibrium properties of statistical systems is nevertheless not feasible. Even for the simplest models, calculations indeed become quickly insurmountable, as in the rod-like chain model when super-structuring becomes dominant~\cite{bouchiat_elastic_2000}. As a result, alternative methods need to be considered. In this regard, {\it bona fide} thermodynamics formalisms, parametrized by the same parameters as those in the underlying statistical system (e.g., supercoiling density), have often proved to be powerful. In particular, approximative solutions can be derived by minimizing the corresponding free energy, which incorporates the interplay between energy and entropy costs that govern the macroscopic behavior of the system. The relevance of these so-called phenomenological approaches lies in the profound connection between statistical mechanics and thermodynamics, where the former provides a microscopic foundation for the latter. The primary challenge then lies in determining a functional form of the free energy that accurately captures the statistical properties of the original system. An illustrative example is the work of Siggia and Marko in 1994, who tackled the issue of super-structuring in a supercoiled DNA chain within the framework of the self-avoiding rod-like chain model. Through the utilization of such a phenomenological approach, they were able to provide an explanation for the higher likelihood of plectonemes (Fig.~\ref{fig:plecto}) compared to toroids, although both types of super-structures may possess the same writhe values~\cite{marko_fluctuations_1994} (section~\ref{sec:3dmodel}).

%\begin{figure}[t]
%\centering
%\includegraphics[width=0.4\linewidth]{./plectoneme.pdf}
%\caption{{\corrprev PLectoneme}}
%\label{fig:plecto}
%\end{figure}

\subsection{Non-equilibrium models}
\label{sec:noneq}

\paragraph*{Out of equilibrium.} While conditions are such that a system is expected to reach thermodynamics equilibrium at long times, the relaxation time required for this equilibrium state to be achieved can be so long that the system may effectively remain out of equilibrium. A prototypical example is the formation of the crumpled (or fractal) globule, a conformation of polymer chains that is predicted to generally occur when the chains are prevented from becoming entangled~\cite{grosberg_crumpled_DNA_1993} -- a situation expected to be relevant for the functioning of cellular DNA. Numerical simulations of high concentrations of polymer chains have revealed that the crumpled globule is a metastable, out of equilibrium conformation that inevitably arises during the swelling of initially condensed, untangled chains, such as those associated with the mitotic chromosomes of eukaryotes~\cite{rosa_interphase_chromosomes_2008,lieberman-aiden_comprehensive_2009} -- see section~\ref{sec:scaleup} for the relevance to the problem of large-scale models of supercoiled DNA. So, while the equilibrium likelihood of the crumpled globule is very low, its lifetime may be so large (hundreds years in the case of the human genome~\cite{rosa_interphase_chromosomes_2008}) that it becomes more relevant than the equilibrium, highly entangled globule.
%With respect to the time scale of the cell cycle, the crumpled globule of the eukaryotic chromatin fiber can thus be considered as a locally equilibrated structure, while being globally out of equilibrium. 
%More generally, in the literature, 
The term ``out of equilibrium'' is then used to describe a situation where thermodynamic concepts, such as free energies, are still applicable in capturing the properties of the corresponding systems. This includes systems that are in the process of reaching thermal equilibration, like the crumpled globule, as well as systems in which perturbations from thermodynamic equilibrium are sufficiently small that their properties can be predicted by considering small deviations from equilibrium statistical mechanics.

\paragraph*{Far from equilibrium.}

In many situations, such as when gyrase utilizes ATP to relax positive supercoils, an additional energy source is required apart from thermal energy. Furthermore, during DNA replication or gene transcription, an influx of matter (nucleotides) is necessary to produce new entities. In these scenarios, describing the system thermodynamically, even in an approximate manner, is often impossible. The associated systems and models are then referred to as being ``far from equilibrium''.
%Numerical simulations (see the next section) of these models, which are frequently employed to investigate equilibrium properties, are necessary to understand the system's characteristics .
Oscillatory systems, which are prevalent in cellular processes, are a prototypical example  that cannot be encompassed within an equilibrium framework since, by definition, they do not satisfy temporal invariance of equilibrium properties.

Interestingly, while gene transcription and DNA replication involve far from equilibrium situations, equilibrium statistical mechanics can still be relevant in explaining certain properties. For example, the general sensitivity of gene transcription to DNA supercoiling can be rationalized using equilibrium-like models~\cite{pineau_what_2022}. This suggests that thermal activation often plays a crucial role in the limiting steps of transcription. A notable example is when the process is primarily limited by the stage where the DNA, bound by the RNAP, must denature to form the open complex. In this case, the rate of transcription is largely determined by the thermodynamic stability of the DNA duplex at the promoter~\cite{forquet_role_2021,forquet_quantitative_2022}.

\subsection{Polymer simulations to investigate the folding of supercoiled DNA}
\label{sec:simus}

Due to the inherent complexity of solving even the simplest model, such as the rod-like chain, numerical simulations of polymer chains are frequently necessary to investigate the folding properties of DNA, whether it is supercoiled or not.
%These simulations have played a crucial role in understanding the mechanisms that govern the cellular organization of DNA in both bacteria and eukaryotes. The significance of numerical simulations stems from the inherent challenge of grasping the spatial properties of self-avoiding (long) polymer chains. 
This section explains the principles of two commonly used methods for conducting such simulations.\\

\paragraph*{The Monte Carlo method.}

One of the most commonly used approaches to simulate an equilibrium situation is the Monte Carlo Markov chain method, often referred to as the Monte Carlo method~\cite{frenkel2001understanding}. The algorithm is relatively straightforward. It involves iterating through a process in which a random entity from the system is selected, and one of its properties is updated with a certain probability. In the context of thermodynamic equilibrium, this probability depends on the change in energy associated with the attempted update and follows a rule known as the ``detailed balance condition''. This condition ensures that, given a sufficiently large number of iterations, the system will reach thermodynamic equilibrium.

For a circular rod-like chain, a typical trial involves randomly selecting two articulation points and rotating the segments located between them by a randomly chosen angle (Fig.~\ref{fig:models}B). This motion, known as a crankshaft motion, alters the bending and torsion energies between the segments at the articulation points and is accepted, or not, following the detailed balance condition. By forbidding rotations that cause segments to intersect in space, the simulated model becomes a self-avoiding rod-like chain.

\paragraph*{The Brownian dynamics method.}

Monte Carlo methods can be efficient in rapidly reaching thermal equilibrium. However, the dynamics they simulate is not realistic, particularly when global moves occur, such as during the rotation of a large number of segments. Therefore, caution must be taken when applying these methods to capture the dynamical properties of real molecules. Similarly, without considering specific types of moves/updates~\cite{liu_efficient_2008}, Monte Carlo ``dynamics'' may not be compatible with certain types of motions that do occur in real molecules, such as the slithering of plectonemes. 
%While additional moves can be considered~\cite{liu_efficient_2008},
To solve these problems and enable a more realistic simulation of the dynamics of DNA, Brownian dynamics simulations~\cite{allison_multistep_1984,chirico_calculating_1992}, which relax the constrain of the rigid segment, are often used. To this end, the DNA chain is described in terms of beads~\cite{chirico_kinetics_1994,jian_combined_1997} (Fig.~\ref{fig:models}C) and its motion is simulated by considering the equations of movement for the beads. Namely, Brownian dynamics simulations assume that DNA beads experience significant frictional forces in the cytoplasm so that their inertia can be neglected. The equations of motion are then numerally solved by updating the positional and rotational degrees considering two main types of forces: (i) those that derive from potential energies, which include the artificial ``bond energy'' between contiguous beads, the bending and torsional energies of the original rod-like chain model, and the short-distance energies of electrostatic repulsion; and (ii) random forces associated with thermal energy, which is responsible for the translational and rotational diffusion of the beads~\cite{chirico_kinetics_1994}. Eventually, by introducing a third category of forces known as active forces~\cite{romanczuk_active_2012}, it becomes possible to simulate far from equilibrium situations, such as the generation of supercoiling due to transcription~\cite{fosado_nonequilibrium_2021,joyeux_models_2022}.

In practice, Brownian dynamics methods allow to simulate in a reasonable amount of time, that is, in less than a few months, $\SI{20}{kb}$ long chains for a total of typically one millisecond~\cite{joyeux_requirements_2020}. This leads to two important remarks. Firstly, similar to Monte Carlo methods, these methods are limited to chains that are typically two orders of magnitude smaller than the typical length of bacterial genomes (megabase pairs). Alternative modeling approaches are thus required to handle larger systems as discussed in section~\ref{sec:scaleup}. Secondly, these methods are not suitable for simulating processes with characteristic time scales on the order of minutes, such as gene transcription or DNA replication. This explains why gene transcription is currently modelled as one-dimensional stochastic processes along the DNA -- see section~\ref{sec:transcription} for more details.}

\begin{acknowledgments}
I.J. would like to thank Thomas Hindré, Gabin Laurent and Olivier Rivoire for their comments on the first version of the review. I.J. and R.E. acknowledge funding from CNRS 80 Prime (MIMIC project), and I.J. and O.E. from ANR-21-CE12-0032 (SISTERS project).
\end{acknowledgments}

\bibliography{bib}

\end{document}